\documentclass[12pt ]{article}
\usepackage{jheppub_kim}
\usepackage{graphicx}
 
\usepackage{graphicx}
\graphicspath{{images/}}
 \usepackage{epsfig}
 
\usepackage{amsmath}
\usepackage{hyperref}
 
\usepackage{float}

\begin{document}

\title{\large Thermodynamics of a newly constructed   black hole coupled with  nonlinear electrodynamics and cloud of strings }

 \author[a]{Himanshu Kumar Sudhanshu,}
	\affiliation[a]{Department of Physics, J. J. College, Magadh University, Gaya, Bihar 823003, India}
\author[b,d]{Dharm Veer Singh,\footnote{Visiting Associate, Inter-University Centre for Astronomy and Astrophysics (IUCAA) Pune-411007, Maharashtra, India}} 
    	 \affiliation[b]{Department of Physics, GLA University, Mathura, Uttar Pradesh 281406, India}
   	
\author[c,d,e]{Sudhaker Upadhyay,\footnote{Visiting Associate, Inter-University Centre for Astronomy and Astrophysics (IUCAA) Pune-411007, Maharashtra, India}}
   	 \affiliation[c]{Department of Physics, K. L. S. College, Nawada, Magadh University, Bodh Gaya, Bihar 805110, India}
   \affiliation[d]{School of Physics, Damghan University, PO Box 3671641167, Damghan,  Iran}
        \author[e]{Yerlan Myrzakulov,}
       \affiliation[e]{Department of General \& Theoretical Physics, L. N. Gumilyov Eurasian National University,  Astana, 010008, Kazakhstan}
       
 \author[e]{and  Kairat Myrzakulov}

 \emailAdd{himanshu4u84@gmail.com} 
   \emailAdd{veerdsingh@gmail.com} 
   \emailAdd{sudhakerupadhyay@gmail.com} 
 \emailAdd{ymyrzakulov@gmail.com}
       \emailAdd{krmyrzakulov@gmail.com}
    
\abstract{ {   This paper finds an exact singular black hole solution in the presence of nonlinear electrodynamics as the source of matter field surrounded by a cloud of strings in $4D$ $AdS$ spacetime. Here, the presence of the cloud of string, the usual Bardeen solution, becomes singular. The obtained black hole solution interpolates with the $AdS$ Letelier black hole in the absence of both the deviation parameter and magnetic charge and interpolates with the $AdS$ Bardeen black hole in the absence of the deviation parameter and a cloud of strings parameter. We analyse the horizon structure and thermodynamics properties, including the stability of the resulting black hole, numerically and graphically. Thermodynamical quantities associated with the black hole get modified due to the nonlinear electrodynamics and cloud of strings.
Moreover, we study the effect of a cloud of strings parameter, magnetic charge and deviation parameter on critical points and phase transition of the obtained black hole where the cosmological constant is treated as the thermodynamics pressure. The critical radius increases with increasing deviation parameter values and magnetic charge values. In contrast, the critical pressure and temperature decrease with increasing deviation parameters and magnetic charge values.
   }}

\maketitle
 
\section{Introduction}\label{sec6.1}

Black holes are one of the most fascinating objects proposed by Einstein's general theory of relativity (EGTR). As a theory of gravitation, in EGTR, gravitational interaction is the consequence of spacetime curvature. Elegant mathematical expressions express the geometric properties of black holes, and hence, obtaining a relevant solution for black holes from EGTR is an intricate and challenging task \cite{stephani2009exact}. The Schwarzschild and Reissner-Nordstr\"{o}m black holes are the initially well-known solutions of the EGTR without matter and with matter in spacetime, respectively. The structure of the spherically symmetric solution often has the singularity \cite{stoica2014geometry}, which is separated by a boundary called the event horizon of a black hole.
On the contrary, Einstein's field equation of general theory of relativity also has non-singular solutions of black holes known as regular black holes, which were first realised by Bardeen, who proposed a regular black hole solution without singularity but has event horizon \cite{bardeen1968proceedings} and interpreted as a solution of Einstein field equation by Ayon-Beato and Gracia in the presence of nonlinear electrodynamics (NLED) \cite{ayon2000bardeen, ayon1999new, ayon1998regular}. Though the Bardeen black hole solution is spherically symmetric, it violates the conditions of strong energy \cite{zaslavskii2010regular}. Later on, many such regular black hole solutions and their properties in spherically symmetric spacetime have been proposed\cite{dymnikova1992vacuum, dymnikova1996sitter, dymnikova2003spherically, hayward2006formation, bronnikov2001regular, bronnikov2017comment, zaslavskii2010regular, rodrigues2018bardeen}. Recently, it has been observed that regular black hole solutions in spherically symmetric spacetime violate the weak energy condition \cite{abbott2021gwtc}. Various solutions of black hole exist in literature in which gravity is coupled with NLED sources, e.g. Born-Infeld black hole \cite{jafarzade2021shadow}, Einstein Gauss-Bonnet black hole \cite{singh2022thermodynamic}, charged AdS black hole \cite{belhaj2020thermal}, charged AdS Einstein Gauss-Bonnet black hole \cite{panah2020charged} and black hole in the presence of cosmic strings \cite{singh2022quasinormal}. It has been discovered that the real linear electromagnetic field in higher energy breaks down due to interaction with other fields. In such conditions, the best possible solution is the interaction of the gravitational field with the NLED. Also, from the perspective of the extension of EGTR, NLED sources of gravity can lead to exciting and challenging geometries, especially in the solution of regular black holes. Hence,  it is essential to focus our attention on finding black hole solutions in NLED and as a consequence of this context, enormous progress has been made \cite{wiltshire1988black, tamaki2000gravitating, breton2003born, fernando2003charged,Singh:2022xgi,Sadeghi:2024krq,Luo:2023ndw,Vishvakarma:2023tnl, Vishvakarma:2023csw, Singh:2021nvm, Singh:2020rnm,Ghosh:2018bxg}.

 In the year 1978, Letelier proposed a new model of a black hole based on pressure-less perfect fluid, an extension of the relativistic "dust cloud" model, known as a cloud of strings (CS)
 \cite{letelier1979clouds, letelier1981fluids, letelier1983string}, which is one-dimensional analogous to a cloud of dust having point particles. However, the whole system with CS is a closed one; its energy-momentum tensor (EMT) is conserved, which leads to applications in cosmology and astrophysics in different situations such as thermodynamics, accretion disk, quasinormal modes and several solutions have been proposed in general relativity and modified theory of gravity context in the presence of CS are reported \cite{ganguly2014accretion, toledo2018black, bronnikov2016birkhoff, toledo2019reissner, mazharimousavi2016cloud, gracca2017quasinormal, barbosa2016rotating, chabab2020thermodynamic, herscovich2010black, cai2020quasinormal, ghosh2014cloud,  Singh:2020nwo}.  {Recently,  a rotating black hole mimicker in the background of the CS is derived, which interpolates to regular black hole spacetime and traversable wormhole \cite{Yang:2023agi}. The gravitational wave echoes of black bounces surrounded by a CS are discussed to explore the effects caused by a CS in the near-horizon region \cite{Yang:2022ryf}. }  Since strings are considered a fundamental component of the universe, this motivates us to investigate black hole solutions in the presence of CS.
 
 Bekenstein and Hawking were the first to realise the black hole thermodynamics by establishing a relation between the entropy and the area of the black hole horizon \cite{Bekenstein:1972tm, PhysRevD.7.2333, PhysRevD.13.191}.  
For leading order, it is sure to the extent that entropy is proportional to the area of the black hole horizon \cite{Strominger:1996sh, Ashtekar:1997yu, Chougule:2023vyb}. 
For higher order, it has been studied widely \cite{Pourhassan:2024yfg, Sudhanshu:2023zle, Kumar:2023agi, Upadhyay:2019hyw, Upadhyay:2017vgk}. It is always interesting to explore 
a new solution for gravity coupled with new sources.
In this connection, this paper provides new regular black hole solutions for Einstein gravity coupled to NLED sources in the presence of CS. Further, we discuss thermodynamics, topology, Joule Thomson effect, stability, $P-v$ criticality and phase transition of this black hole \cite{Zhang:2023, Zhang:2023tlq, Zhang:2024fxj, Chen:2024kmy, Chen:2024atr}.
  
 The plan and structure of this paper are as follows. In section \ref{sec6.2}, we consider the Einstein gravity coupled to NLED in $4D$ AdS spacetime to obtain an exact black hole solution in the presence of the CS and further discuss its horizon structure for different values of parameters. We discuss the thermodynamics quantities, stability, and phase transitions in section \ref{sec6.3}. In section \ref{sec6.4}, we report this black hole's $P-v$ criticality and phase transitions analogous to the Van der Waals (VdW) fluid. We have calculated the critical values of horizon radius (or specific volume), critical pressure, critical temperature, and the universal critical compressibility factor, as well as their dependency on various parameters. Finally, in the last section, we discuss the results and final remarks on this new black hole solution and its thermodynamic properties. 

For all our calculations, we adopt $(-, +, +, +)$ as metric signature and work in natural units where $G = \hbar = c = \kappa_B = 1$. 

\section{A New Regular Black Hole Solution}\label{sec6.2}
The Einstein-Hilbert action describing Einstein's gravity coupled to NLED sources surrounded by a CS in $4D$ AdS spacetime \cite{ayon2000bardeen, singh2022quasinormal} is given by

\begin{equation}
{\cal S}= \int d^{4}x\sqrt{-{\tilde{g}}}\left[\frac{1}{2 \kappa}\left(R-2\Lambda \right) -\frac{1}{4 \pi} {\cal L}^{NE}(F)\right] +S^{CS},
\label{NLEDaction}
\end{equation}
where $\tilde{g}$ is determinant of metric,  $R$ is Ricci curvature scalar and $\Lambda=-\frac{3}{l^2}$ is a cosmological constant in which $l$ is the AdS length. ${\cal L}^{NE}(F)$  and $S^{CS}$ describe Lagrangian density for NLED sources and action for CS sources, respectively, which are specified below. The Lagrangian density of NLED sources, ${\cal L}^{NE}(F)$ is function of $F=F_{\mu \nu} F^{\mu \nu}/4$, where $F_{\mu \nu}$ if the electromagnetic field strength tensor which is associated with the gauge potential $A_\mu$ as $F_{\mu \nu}=2\nabla_{[\mu} A_{\nu]}$.

The ${\cal L}^{NE}(F)$  is the Lagrangian density of the NLED  satisfying the weak energy condition ($F<<1$),  which must be a continuous function of $F$   such that $\partial L(F)^{NE}/\partial F\to \infty$ as $F\to\infty$ and $ {\cal L}^{NE}_F\to 1$ as $F=0$. Keeping this in mind, along with the regularity of spacetime, we choose
\begin{equation}
{\cal L}^{NE}(F)= \frac{F e^{-s(2g^2F)^{1/4}}}{1+(2g^2F)^{3/4}}\left[1+\frac{3}{s}\left(\frac{(2g^2F)^{1/2}}{1+(2g^2F)^{3/4}}\right)\right],
\label{NLEDsources}
\end{equation}
where $s$ is defined in terms of free parameters $g$ and $M$ corresponds to the magnetic charge and mass of the black hole, respectively, as $s=|g|/2M$. The Lagrangian density (${\cal L}^{NE}(F)$) in the limit of $F<<1$ is
\begin{eqnarray}
  {\cal L}^{NE}= F+F^{\frac{3}{2}}\left(\frac{3g}{2\sqrt{2}}+\frac{s^2 g}{\sqrt{2}}\right)-sF^{\frac{5}{4}}(2g^2)^{\frac{1}{4}} +F^{\frac{7}{4}}\left((2g^2)^{\frac{3}{4}}-\frac{3s g^{\frac{3}{2}}}{2^{\frac{5}{4}}}-\frac{s^3g^{\frac{3}{2}}}{3 2^{\frac{1}{4}}}\right)+\hdots.
  \label{lw}
\end{eqnarray}
In the weak field limit the Lagrangian density ($L(F)^{NE}$) goes over $L(F)^{NE}=F$,  whereas in
the strong-field limits, it vanishes. Here, we must emphasise that for gravity coupled to NLED with Lagrangian density, $L(F)^{NE}$, there exist spherically symmetric solutions having globally regular metric which possesses a correct weak field limit for the magnetic case only. However, one can explain the electric analogues of magnetic solutions with different Lagrangian densities (using Legendre transformation in the Hamiltonian formalism) for various ranges of radial coordinates.
 
 The cloud of strings source is governed by the Nambu-Goto action, which is used to describe strings like object \cite{letelier1979clouds, letelier1983string} and is given by 
\begin{equation}
S^{CS}=\int \sqrt{-\gamma}{\cal M}d\lambda^0 d\lambda^1=\int {\cal M} \left(-\frac{1}{2} \Sigma^{\mu \nu}\Sigma_{\mu \nu}\right)^{\frac{1}{2}}d\lambda^0 d\lambda^1, 
\label{ngaction}
\end{equation}
where $\cal M$ is the dimensionless constant which characterizes the string, $(\lambda^0, \lambda^1)$ are the timelike and spacelike coordinate parameters, respectively \cite{synge1960relativity}. $\gamma$ is the determinant of the induced metric, $\gamma_{a b}=\tilde{g}_{\mu \nu}\frac{\partial x^{\mu}}{\partial \lambda^a}\frac{\partial x^{\nu}}{\partial \lambda^b}$ of the strings world sheet. $\Sigma^{\mu \nu}=\epsilon^{ab}\frac{\partial x^{\mu}}{\partial \lambda^a}\frac{\partial x^{\nu}}{\partial \lambda^b}$ is bivector related to string world sheet, where $\epsilon^{ab}$ is the second rank Levi-Civita tensor which takes the non-zero values as $\epsilon^{01}=-\epsilon^{10}=1$.

Varying the action (\ref{NLEDaction}) with respect to the metric, $\tilde{g}_{\mu \nu}$ and the gauge potential, $A_{\mu}$, we can obtained the equations of motion as

\begin{eqnarray}
&&G_{\mu \nu }+\Lambda \tilde{g}_{\mu \nu }=T_{\mu\nu}^{NE}+T_{\mu\nu}^{CS},
\label{nledeom1}\\
   && \nabla_{\mu}\left(\frac{\partial {\cal L}^{NE}(F)}{\partial F}F^{\mu \nu}\right)=0,\qquad\text{and} \qquad \nabla_{\mu}(* F^{\mu\nu})=0,
   \label{nledeom2}
\end{eqnarray}
where $G_{\mu \nu}$ is the Einstein tensor, $T_{\mu\nu}^{NE}$ and $T_{\mu\nu}^{CS}$ are EMT associated with NLED and CS sources,  respectively.

The EMT related to the NLED source (\ref{NLEDsources}) is given by

\begin{equation}
T_{\mu\nu}^{NE}= 2\left[\frac{\partial{\cal L}^{NE}(F)}{\partial F}F_{\mu \sigma}F^{\sigma}_{\nu}-\tilde{g}_{\mu \nu} {\cal L}^{NE}(F)\right],
\label{nledtensor}
\end{equation}

Now, to find the black hole solution with NLED sources in the presence of CS, one can consider the static spherically symmetric spacetime line element  $(\kappa=c=1)$ in $4D$ spacetime as

\begin{equation}
ds^2=-f(r)dt^2 +\frac{1}{f(r)}dr^2+r^2 d\Omega^2, \quad \text{with} \quad f(r)=1-\frac{2m(r)}{r}, 
\label{nledlemt}
\end{equation}
where $d\Omega^2 = d\theta^2+\sin^2 \theta d\phi^2$ is metric of unit $2D$ sphere. The metric function $f(r)$ is to be determined by solving the field equations (\ref{nledeom1}) and (\ref{nledeom2}).

{To determine the metric function, we examine the following magnetic charge choice for Maxwell's field strength tensor $F_{\mu\nu}$
\begin{equation}
  F_{\mu\nu}=2\delta^{\theta}_{[\mu}\delta^{\phi}_{\nu]} Z(r,\theta).
  \label{fab}
\end{equation}
Substituting ansatz (\ref{fab}) in equation of motion (\ref{nledeom2}) and integrating, we obtain \cite{ayon2000bardeen}
 \begin{equation}
F_{\mu\nu}=2\delta^{\theta}_{[\mu}\delta^{\phi}_{\nu]}h(r) \sin\theta.
\label{emt2}
\end{equation}
The non-vanishing component of $F_{\mu\nu}$ is $F_{\theta\phi}=h(r)\sin\theta$ with potential $A_\phi=-h(r)\cos\theta$ \cite{ayon2000bardeen}. Using $dF=0$; Hence $g(r)\sin\theta dr\wedge d\theta\wedge d\phi=0$, which conclude that $h(r)= \text{constant}=g$. Here, $g$ is the magnetic charge. Hence, the magnetic field strength is given by 
\begin{equation}
F_{\theta\phi}= g\sin\theta \qquad\text{and} \qquad F=\frac{1}{2}\frac{g^2}{r^4}.
\label{emt3} 
\end{equation}}


Now, substituting the value of magnetic field strength, $F$ from equation (\ref{emt3}) to equation (\ref{NLEDsources}), one can obtain the Lagrangian density of NLED sources as

\begin{equation}
{\cal L}^{NE}(F)=\frac{g^2}{2r}\frac{e^{-sg/r}}{(r^3+g^3)}\left[1+\frac{3}{s}\left(\frac{g^2r}{r^3+g^3}\right)\right],
\label{emt1}
\end{equation}
and using equation (\ref{nledtensor}) and equation (\ref{emt3}), the ${T^{NE}}_{t}^{t}$ and $ {T^{NE}}_{r}^{r}$ components of energy momentum tensor is obtained as

\begin{equation}
{T^{NE}}_{t}^{t}= {T^{NE}}_{r}^{r}= -2{\cal L}^{NE}(F)= -\frac{2M e^{-k/r}}{(r^3+g^3)}\left(\frac{k}{r}+\frac{3 g^3}{r^3+g^3}\right),
\label{nledemtcomp}
\end{equation}
where $k=g^2/2M$ is known as deviation parameter and the equation (\ref{nledemtcomp}) satisfy the equation of motion for NLED.

 The EMT for the CS (\ref{ngaction}) is calculated from the definition as
\begin{equation}
T_{\mu\nu}^{CS}=2 \frac{\partial}{\partial {\tilde{g}}_{\mu \nu}}{\cal M}\left(-\frac{1}{2}\Sigma^{\mu \nu}\Sigma_{\mu \nu}\right)^{1/2}= \frac{\rho \Sigma_{\alpha \nu} \Sigma_{\mu}^{\alpha}}{\sqrt{-\gamma}},
\label{CSTensor}
\end{equation}
where $\rho$ is the proper density of the CS. From the conservation of law, $\nabla_{\mu}T_{\mu \nu}^{CS}=0$, we obtain the non-vanishing components of the EMT of the CS as
\begin{equation}
{T^{CS}}^t_t= {T^{CS}}^r_r= -\frac{a}{r^2},
\label{csemtcomp}
\end{equation}
where $a$ is the constant known as the CS parameter.

Using the Eq. (\ref{nledlemt}) the value of $(r,r)$ components in equation (\ref{nledeom1}), one get, 
\begin{equation}
m'(r)=-\frac{3r^2}{2l^2}+\frac{a}{2}-\frac{M r e^{-k/r}}{(r^3+g^3)}\left(k+\frac{3g^3 r}{r^3+g^3}\right).
\label{rreq1}
\end{equation}
On integrating  equation (\ref{rreq1}) with respect to $r$ from $r$ to $\infty$, we get
\begin{equation}
\label{intmr}
m(r)=-\frac{r^3}{2l^2}+\frac{ar}{2}-\int_r^{\infty}\left[\frac{M r e^{-k/r}}{(r^3+g^3)}\left(k+\frac{3 g^3 r}{r^3+g^3}\right)\right]dr +C_1.
\end{equation}
Here, $C_1$ is the integration constant determined through condition $\lim_{r\to\infty} \left(m(r)+\frac{r^3}{2l^2}- \frac{ar}{2}\right)=M$  (mass of black hole) such that
\begin{equation}
\label{intconst}
C_1= M,
\end{equation}
and
\begin{equation}
\int_r^{\infty}\left[\frac{M r e^{-k/r}}{(r^3+g^3)}\left(k+\frac{3 g^3 r}{r^3+g^3}\right)\right]dr=M-\frac{Mr^3}{r^3+g^3}e^{-k/r}.
\label{1}
\end{equation}

Substituting the value of (\ref{intconst}) and (\ref{1}) into Eq. (\ref{intmr}), we get
\begin{equation}
m(r)=\frac{M r^3}{r^3+g^3}e^{-k/r}-\frac{r^3}{2l^2}+\frac{ra}{2}.
\label{2}
\end{equation} 
and the metric function for a $4D$ black hole with NLED sources in the presence of CS is obtained as
\begin{equation}
f(r) =1-a+\frac{r^2}{l^2}-\frac{2Mr^2e^{-k/r}}{r^3+g^3}.
\label{bhsnled}
\end{equation}

The obtained regular black hole solution (\ref{bhsnled}) is characterised by mass, $M$, cosmological constant as $l=\sqrt{-3/\Lambda}$, the cloud of the string parameter, $a$, magnetic charge, $g$ and the deviation parameter, $k$, which ensure its deviation from the Bardeen black hole. For $k = g = 0$, it reduces to the Letelier solution \cite{{letelier1979clouds}}, and also for $a = 0$, it corresponds to the Schwarzschild black hole solution in $AdS$ spacetime. In the limit $r<<1$, the obtained black hole solution (\ref{bhsnled}) becomes
\begin{equation}
   f(r)=1-a-\frac{r^2}{l^2_{eff}}+\frac{2Mk}{g^3} 
   \label{3}
\end{equation}
where $1/l^2_{eff}=2M/g^3-1/l^2$ and in the limit of $r>>1$ is
\begin{equation}
    f(r)=1-a-\frac{2M}{r}+\frac{r^2}{l^2}
    \label{4}
\end{equation}
 To study the nature of singularities structure of this black hole (\ref{bhsnled}) at $r=0$, it becomes essential to analyse the curvature invariants of the spacetime such as Ricci scalar, $R$, Ricci square, $R_{\mu\nu}R^{\mu\nu}$ and Kretshmann scalars, $R_{\mu\nu\rho\sigma}R^{\mu\nu\rho\sigma}$.
 
\begin{eqnarray}
&&R =-\frac{12}{l^2}+\frac{2a}{r^2}-\frac{12 M r^2 e^{-k/r}}{A^2}\left(k+5r-\frac{3r^4}{A}\right)+\frac{2M e^{-k/r} }{A}\left(12+\frac{6k}{r}+\frac{k^2}{r^2}\right),\\
 &&R_{\mu\nu}R^{\mu\nu}=\frac{8a M e^{-\frac{k}{r}}}{A}\left(\frac{3}{r^2}+\frac{k}{r^3}-\frac{3r}{A}\right)-\frac{12 M e^{-\frac{k}{r}}}{l^2 A} \left(12 -\frac{6r^2(k+5)}{A}+\frac{k(1+6r)}{r^2}+\frac{18 r^6}{A^2}\right)\nonumber\\&&\qquad+\frac{2a(al^2-6r^2)}{l^2r^4}+\frac{2 M^2 e^{-2\frac{k}{r}}}{A^2} \left(\frac{324 r^{12}}{A^4}-\frac{216r^8(k-r)}{A^3}+\frac{36(2k^2r^4+23r^6-12kr^4)}{A^2}\right)\nonumber\\&&\qquad+\frac{36}{l^4}-\frac{12}{A}\left(k^3+8k^2 r+24 k r^2-30 r^3\right)+\left(\frac{72(k+r)}{r}+\frac{32 k^2}{r^2}+\frac{8k^3}{r^3}+\frac{k^4}{r^4}\right),\\
 &&R_{\mu\nu\rho\sigma}R^{\mu\nu\rho\sigma} =\frac{12 M e^{-k/r}}{l^2A}\left(\frac{24 r^6}{A^2}-\frac{1}{A}\left(8kr^2+28r^3-3r^5\right)+\left(4+k r-3r^2\frac{4k^2}{3r^2}+\frac{4k}{r}\right)\right)\nonumber\\&&\qquad+\frac{9r^2}{l^4}-\frac{12}{l^2}+a^2\left(\frac{4}{r^4}+\frac{1}{r^2}\right)-2a\left(\frac{4}{l^2r^2}+\frac{3}{l^2}\right)+\frac{4 M^2 e^{-2k/r}}{A^2} \left(\frac{216}{A^3}\right.\nonumber\\&&\qquad- \frac{1}{A^2}(144+360r^6+9r^8)+\frac{1}{A}(36 r +144kr^2+168r^3-6k-18)+6 a r-12\nonumber\\
 &&\qquad-\frac{16k^2}{r^2}-\frac{24}{r}+k^2-4k^3-16k^2+9r^2\Big),
\end{eqnarray}
 where $A=g^3+r^3$. Here, we see that these curvature invariants diverge in the limit $r\rightarrow 0$ in the presence of the CS parameter. This result matches with the result shown in \cite{rodrigues2022embedding}. Hence, this black hole solution (\ref{bhsnled}) is singular everywhere in spacetime and hence, the obtained black hole solution (\ref{bhsnled}) is singular. But in the absence of the CS parameter, the invariants are well-behaved with finite constant values at the origin, indicating the regular solution.

 \begin{figure}[]
\begin{tabular}{c c} 
\includegraphics[width=.5\linewidth]{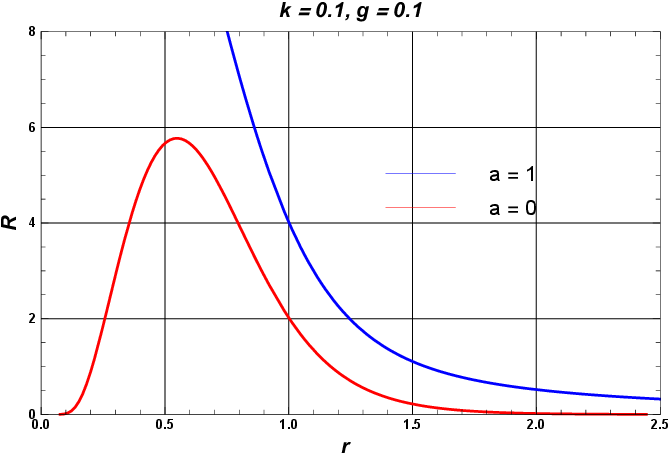} 
\includegraphics[width=.5\linewidth]{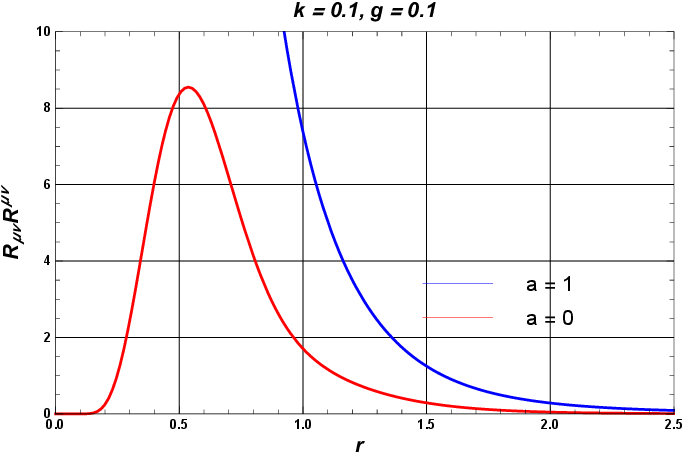}\\
\includegraphics[width=.5\linewidth]{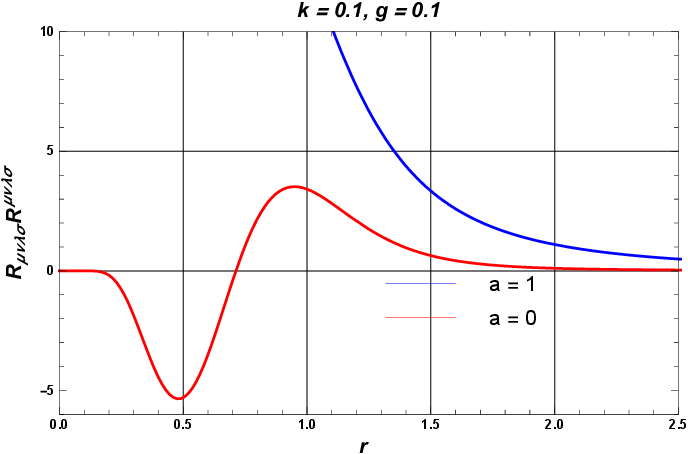}
\end{tabular}
\caption{The Plot of curvature invariants vs radial distance with a fixed value of $k=0.1$, $g=0.1$ and different $a$ with $M=1$ and $l=10$.}
\label{fig:fr} 
\end{figure}

{ Now, the nature of the black hole horizon for the solution obtained in (\ref{bhsnled}) can be determined by equations $f(r)=0$ and $f' (r)=0$ \cite{Rodrigues:2022zph}. We can obtain the horizon of the obtained black hole solution and the degenerate horizon}. The equation (\ref{bhsnled}) is transcendental. Hence its analytic solution does not exist and thus can be solved numerically only as shown in Fig. \ref{fig:fr} and its numerical values of the Cauchy (or inner) horizon $(r_{-})$ and the event (or outer) horizon $(r_{+})$ along with their deviation $\delta(=r_+ - r_-)$  are shown in Table \ref{tab:fr} for different values of parameters.
 
\begin{figure}[!]
\begin{tabular}{c c} 
\includegraphics[width=.50\linewidth]{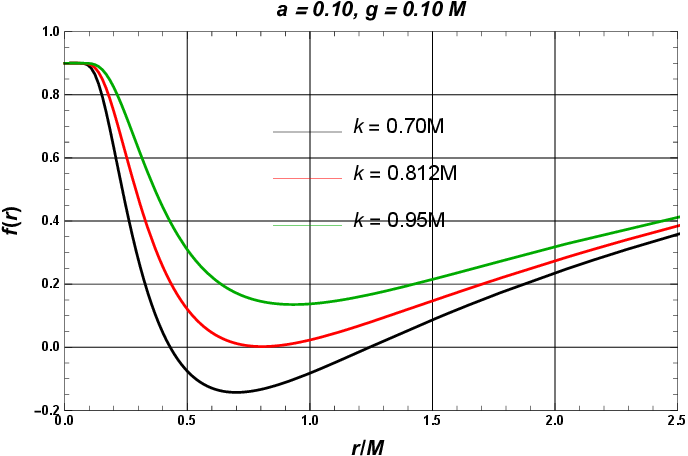} 
\includegraphics[width=.50\linewidth]{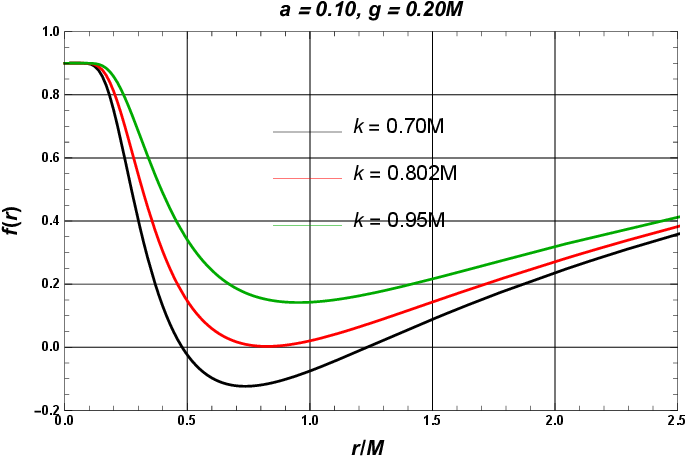}\\
\includegraphics[width=.50\linewidth]{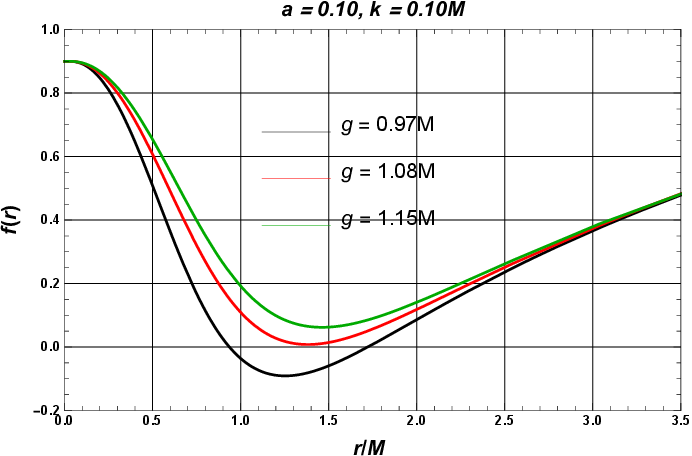}
\includegraphics[width=.50\linewidth]{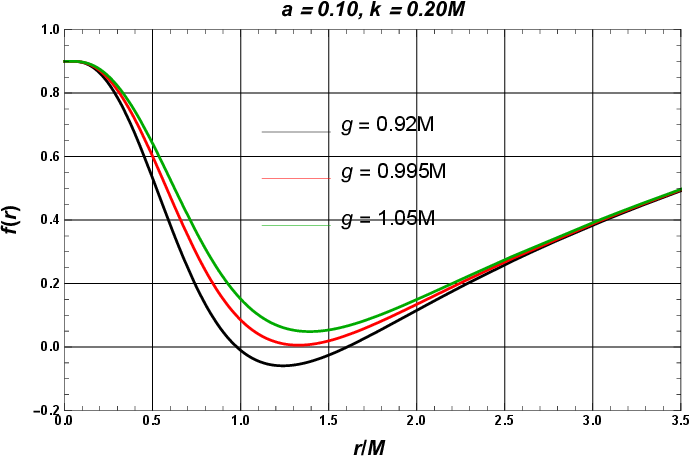}
\end{tabular}
\caption{The Plot of metric function, $f(r)$ versus horizon radius, $r$ of $4D$ NLED black hole in a CS for different value of $k$, $g$ and $a$ with $M=1$ and $l=10$.}
\label{fig:fr} 
\end{figure}

As we notice from Fig. \ref{fig:fr} and Table \ref{tab:fr}, the horizon structure of the black hole depends on the parameters $a, g$ and $k$. There is no horizon radius for a value of deviation parameters $k > k_c$ or magnetic charge $g > g_c$, which is called the critical value. There exists both Cauchy horizon $(r_{-})$ and event horizon $(r_{+})$ for parameters $k$ or $g$ less than its critical value. For the critical value of these parameters, only the event horizon of the black hole exists. It is worth mentioning that the horizon of the black hole decreases with the increase in the value of $g$ and $k$ but increases with the increase in  $a$. In addition, the critical value of the magnetic charge, $g_{c}$ or the deviation parameter, $k_{c}$, increases with increasing the value of $a$.

\begin{table}[h!tb]
\centering 
\begin{tabular}{cccccccc}
\hline
\multicolumn{4}{c|}{$a   = 0.10$, $g = 0.10$}                                                        & \multicolumn{4}{c}{$a =   0.20$, $g = 0.10$}                                                \\ \hline \hline
$k$                  & $r_-$                & $r_+$                & \multicolumn{1}{c|}{$\delta$} & $k$                  & $r_-$                & $r_+$                & $\delta$             \\ \hline
0.20                  & 0.087                & 1.923                & \multicolumn{1}{c|}{1.837}    & 0.20                  & 0.083                & 2.153                & 0.901                \\
0.40                  & 0.167                & 1.702                & \multicolumn{1}{c|}{0.619}    & 0.40                  & 0.158                & 1.943                & 1.785                \\
0.60                  & 0.310                & 1.426                & \multicolumn{1}{c|}{1.116}    & 0.60                  & 0.280                & 1.693                & 1.413                \\
0.80                  & 0.682                & 0.930                & \multicolumn{1}{c|}{0.248}    & 0.80                  & 0.501                & 1.352                & 0.851                \\ \hline
\multicolumn{2}{c}{$k_c$ =0.810}            & 0.803                & \multicolumn{1}{c|}{}         & \multicolumn{2}{c}{$k_c$ =0.909}            & 0.895                &                      \\ \hline \hline
\multicolumn{1}{l}{} & \multicolumn{1}{l}{} & \multicolumn{1}{l}{} & \multicolumn{1}{l}{}          & \multicolumn{1}{l}{} & \multicolumn{1}{l}{} & \multicolumn{1}{l}{} & \multicolumn{1}{l}{} \\ \hline 
\multicolumn{4}{c|}{$a = 0.10$, $k = 0.10$}                                                          & \multicolumn{4}{c}{$a = 0.20$, $k = 0.10$}                                                  \\ \hline \hline
$g$                  & $r_-$                & $r_+$                & \multicolumn{1}{c|}{$\delta$} & $g$                  & $r_-$                & $r_+$                & $\delta$            \\ \hline
0.20                  & 0.104                & 2.021                & \multicolumn{1}{c|}{1.917}    & 0.20                  & 0.099                & 2.248                & 2.148                \\
0.40                  & 0.230                & 2.008                & \multicolumn{1}{c|}{1.778}    & 0.40                  & 0.217                & 2.237                & 2.020                \\
0.60                  & 0.403                & 1.969                & \multicolumn{1}{c|}{1.566}    & 0.60                  & 0.375                & 2.208                & 1.833                \\
0.80                  & 0.639                & 1.883                & \multicolumn{1}{c|}{1.244}    & 0.80                  & 0.581                & 2.145                & 1.564                \\ \hline
\multicolumn{2}{c}{$g_c$ = 1.070}           & 1.371                & \multicolumn{1}{c|}{}         & \multicolumn{2}{c}{$g_c$ = 1.203}           & 1.524                &                      \\ \hline \hline
\end{tabular} 
\caption[Numerical data of Cauchy and event horizons of $4D$ NLED black hole in CS for different values of parameters $a, g$ and $k$]{For different values of parameters $a$, $g$ and $k$ with $M=1$ and $l=10$, Cauchy horizon $(r_-)$, event horizon $(r_+)$ and their deviation $\delta(=r_+ - r_-)$ has been tabulated numerically for $4D$ NLED black hole in CS.}
\label{tab:fr} 
\end{table}

\section{Thermodynamics of NLED Black Hole in CS} \label{sec6.3}
In this section, we study the thermodynamic properties of a $4D$ AdS black hole with NLED sources in a CS. In this connection, we discuss the thermodynamic parameters of the mass, temperature, entropy, specific heat capacity and Gibbs free energy at the event horizon for different values of parameters $a$, $g$ and $k$ of the black hole \cite{Rodrigues:2022zph, Rodrigues:2022qdp}.  

\begin{figure}[!]
\begin{tabular}{c c} 
\includegraphics[width=.50\linewidth]{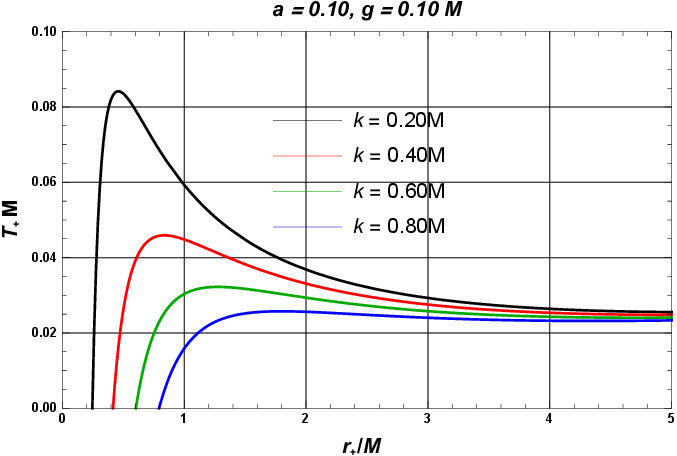} 
\includegraphics[width=.50\linewidth]{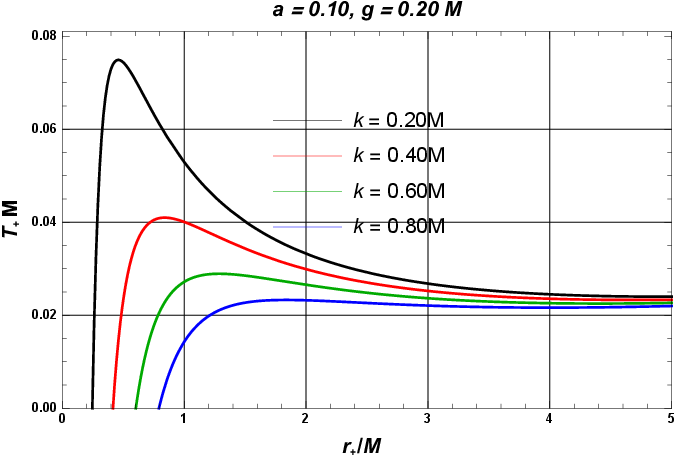}\\
\includegraphics[width=.50\linewidth]{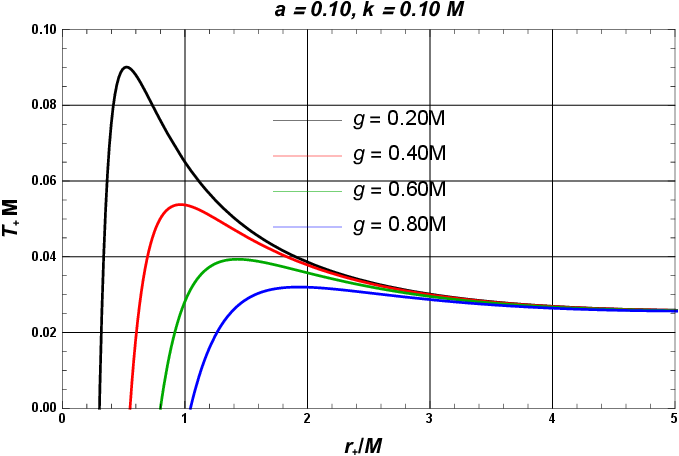}
\includegraphics[width=.50\linewidth]{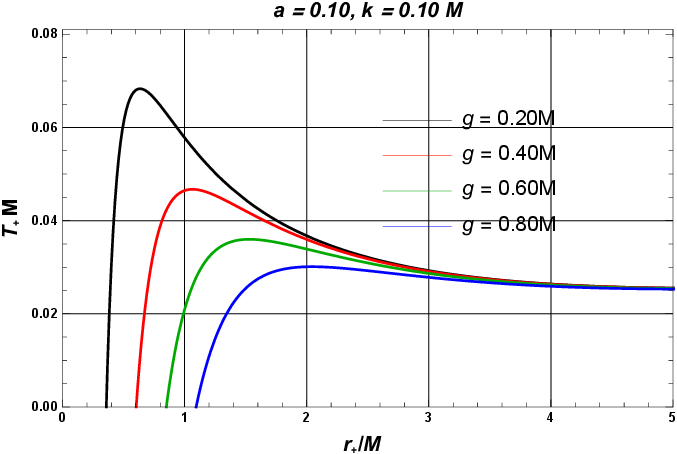}
\end{tabular}
\caption[Temperature plot of $4D$ NLED black hole in a CS]{Plot of Hawking temperature $T_+$ versus horizon radius, $r_+$ of $4D$ NLED black hole in a CS for different value of $k$, $g$ and $a$ with $l=10$.}
\label{fig:t} 
\end{figure}

The mass of the black hole, $M$ can be evaluated from metric function (\ref{bhsnled}) at the horizon by setting metric function $f(r)|_{r=r_+}=0$ as
\begin{equation}
M=\frac{\left(r_+^3+g^3\right)  \left(1-a +\frac{l^2}{r_+^2}\right)}{2 r_+^2}e^{k/r_+}.
\label{eqn:m}
\end{equation}
The expression represents the mass of a $4D$ AdS black hole with NLED in the CS. { The mass of the obtained black hole solution (\ref{bhsnled}) reduces to the mass of the Letelier black hole with CS in the absence of magnetic monopole charge, $AdS$ Hayward black hole in the absence of ($k,a$) \cite{Luo:2023ndw, Sadeghi:2024krq} and $AdS$ regular black hole in the absence of ($g,a$) \cite{Singh:2022xgi}.} For various values of parameters $a, g$, and $k$, the mass $(M)$ has been plotted with horizon radius $(r_+)$. As the radius of the black hole decreases, its mass decreases linearly, like the Schwarzschild black hole. At a particular critical horizon radius, its mass decreases to a minimum value, after which mass increases exponentially and tends to infinity (singularity) as the horizon radius approaches zero.
Further, we observe that the black hole's mass increases with the value of   $g$ or   $k$. At the same time, it decreases with an increase in the value of the CS parameter $(a)$. It reduces to the mass of $4D$ $AdS$ Schwarzschild black hole for vanishing the value of all the three parameters $a, g$ and $k$. The minimum value of mass and its critical radius increases with the values of  $k$ and  $g$. In contrast, it has the opposite variation for the CS parameter, $a$. 
The Hawking temperature for the regular black hole solution (\ref{bhsnled}) can be estimated from the standard formula
\begin{equation}
T_+=\frac{1}{4\pi}f'(r)|_{r=r_{+}},
\end{equation} 
where $f'(r)$ denotes the differentiation of metric function (\ref{bhsnled}) with respect to $r$, and in our case it is obtained as 
\begin{equation}
T_+ =\frac{1}{4 \pi l^2 r_+^2}\left[2 r_+^3-\frac{\left(r_+^2-(a-1) l^2\right) \left(r_+^3 (k-r_+)+g^3 (k+2 r_+)\right)}{(r_+^3+g^3)}\right].
\label{eqn:t}
\end{equation}
The Hawking temperature, $T_+$ (\ref{eqn:t}), is characterized by the black hole parameters $a$, $g$ and $k$ and on cosmological length, $l$. It is clear from the expression that it reduces to the temperature of $4D$ AdS Schwarzschild black hole in the limit all the three parameters $a, g$ and $k$ tends to zero and it reduces to 
\begin{equation}
\label{tempschw2}
T_+^{Swzc}=\frac{1}{4 \pi}\left(\frac{1}{r_+}+\frac{3}{l^2}\right).
\end{equation}
We plot the temperature (\ref{eqn:t}) of the $4D$ AdS black hole with NLED in the CS in Fig. \ref{fig:t} for different values of $g$, $k$ and CS parameters $(a)$ to analyse the functioning of temperature with the horizon radius of this black hole. { The temperature of the obtained black hole solution (\ref{bhsnled})  reduces to the temperature of the Letelier black hole with CS in the absence of magnetic monopole charge, $AdS$ Hayward black hole in the absence of ($k,a$) \cite{Luo:2023ndw,Sadeghi:2024krq} and $AdS$ regular black hole in the absence of ($g,a$) \cite{Singh:2022xgi}.}
From the plot, we observe that the effect of $g$, $k$ or $a$ is more significant in the case of small black holes, and for the larger horizon region, it coincides with that of Schwarzschild black hole. The temperature of this regular black hole first increases sharply to attain a maximum value, $T_+^{max}$ for a particular horizon radius, $r_+^m$ and then decreases exponentially with an increase in horizon radius, $r_+$ and finally, in the region of larger black holes, it corresponds to temperature (\ref{tempschw2}) of $4D$ AdS Schwarzschild black hole $(T_+^{Swzc})$. The maximum of Hawking temperature, $T_+^{max}$ 
and its corresponding horizon radius, $r_+^m$ have been computed numerically for the different values of $g$, $k$ and CS parameters $(a)$ as shown in Table \ref{tab:tmax}. Here, we observe that the value of maximum temperature, $T_+^{max}$, increases and its corresponding horizon radius, $r_+^m$, decreases with a decrease in the value of parameters $a$, $g$ or $k$.

\begin{table}[h!tb]
\centering 
\begin{tabular}{ccccccccc}
\hline
\multicolumn{9}{c}{$a = 0.10$}                                                                                                                                                                                                                        \\ \hline \hline
\multicolumn{1}{c|}{}    & \multicolumn{2}{c|}{$g = 0.10$}                          & \multicolumn{2}{c|}{$g = 0.20$}                          & \multicolumn{2}{c|}{$g = 0.30$}                          & \multicolumn{2}{c}{$g = 0.40$}               \\ \hline
\multicolumn{1}{c|}{$k$} & $r_+^{m}$            & \multicolumn{1}{c|}{$T_+^{max}$} & $r_+^{m}$            & \multicolumn{1}{c|}{$T_+^{max}$} & $r_+^{m}$            & \multicolumn{1}{c|}{$T_+^{max}$} & $r_+^{m}$            & $T_+^{max}$          \\ \hline
\multicolumn{1}{c|}{0.10} & 0.315                & \multicolumn{1}{c|}{0.135}       & 0.524                & \multicolumn{1}{c|}{0.090}       & 0.742                & \multicolumn{1}{c|}{0.067}       & 0.965                & 0.054                \\
\multicolumn{1}{c|}{0.20} & 0.459                & \multicolumn{1}{c|}{0.084}       & 0.635                & \multicolumn{1}{c|}{0.068}       & 0.843                & \multicolumn{1}{c|}{0.056}       & 1.064                & 0.047                \\
\multicolumn{1}{c|}{0.30} & 0.638                & \multicolumn{1}{c|}{0.059}       & 0.772                & \multicolumn{1}{c|}{0.054}       & 0.963                & \multicolumn{1}{c|}{0.047}       & 1.177                & 0.041                \\
\multicolumn{1}{c|}{0.40} & 0.837                & \multicolumn{1}{c|}{0.046}       & 0.936                & \multicolumn{1}{c|}{0.043}       & 1.103                & \multicolumn{1}{c|}{0.040}       & 1.307                & 0.036                \\ \hline \hline
\multicolumn{1}{l}{}     & \multicolumn{1}{l}{} & \multicolumn{1}{l}{}             & \multicolumn{1}{l}{} & \multicolumn{1}{l}{}             & \multicolumn{1}{l}{} & \multicolumn{1}{l}{}             & \multicolumn{1}{l}{} & \multicolumn{1}{l}{} \\ \hline 
\multicolumn{9}{c}{$a = 0.20$}                                                                                                                                                                                                                        \\ \hline \hline
\multicolumn{1}{c|}{}    & \multicolumn{2}{c|}{$g = 0.10$}                          & \multicolumn{2}{c|}{$g = 0.20$}                          & \multicolumn{2}{c|}{$g = 0.30$}                          & \multicolumn{2}{c}{$g = 0.40$}               \\ \hline
\multicolumn{1}{c|}{$k$} & $r_+^{m}$            & \multicolumn{1}{c|}{$T_+^{max}$} & $r_+^{m}$            & \multicolumn{1}{c|}{$T_+^{max}$} & $r_+^{m}$            & \multicolumn{1}{c|}{$T_+^{max}$} & $r_+^{m}$            & $T_+^{max}$          \\ \hline
\multicolumn{1}{c|}{0.10} & 0.315                & \multicolumn{1}{c|}{0.120}       & 0.524                & \multicolumn{1}{c|}{0.080}       & 0.743                & \multicolumn{1}{c|}{0.060}       & 0.967                & 0.048                \\
\multicolumn{1}{c|}{0.20} & 0.460                & \multicolumn{1}{c|}{0.075}       & 0.635                & \multicolumn{1}{c|}{0.061}       & 0.845                & \multicolumn{1}{c|}{0.050}       & 1.067                & 0.042                \\
\multicolumn{1}{c|}{0.30} & 0.639                & \multicolumn{1}{c|}{0.053}       & 0.774                & \multicolumn{1}{c|}{0.048}       & 0.966                & \multicolumn{1}{c|}{0.042}       & 1.181                & 0.037                \\
\multicolumn{1}{c|}{0.40} & 0.839                & \multicolumn{1}{c|}{0.041}       & 0.939                & \multicolumn{1}{c|}{0.039}       & 1.108                & \multicolumn{1}{c|}{0.036}       & 1.313                & 0.032                \\ \hline \hline
\end{tabular} 
\caption[Numerical data of maximum temperature and its corresponding horizon radius of $4D$ NLED black hole in CS for different values of parameters $a, g$ and $k$]{For different values of parameters $a, g$ and $k$ with $l=10$, maximum value of Hawking temperature, $T_+^{max}$ and its corresponding horizon radius, $r_+^m$  have been tabulated numerically for $4D$ NLED black hole in CS.}
\label{tab:tmax} 
\end{table}

Now, let's focus on the other most crucial thermodynamic quantity, the entropy of the black hole, by considering the first law of thermodynamics.
Being a thermodynamic system, the black hole must follow the first law of thermodynamics, defined as 

\begin{equation}
dM=T_+dS_+ + \Phi_{g} dg + V_+dP_+,
\label{fst}
\end{equation} 
where $S_+$ is the entropy of the black hole and $\Phi_{g}$ is the potential for the magnetic charge, $g$. The thermodynamic pressure, $P_+$ and its conjugate thermodynamic volume, $V_+$, of the black hole is given as

\begin{equation}
P_+=- \frac{\Lambda}{8 \pi} = \frac{3}{8 \pi l^2},
\label{eqn:p}
\end{equation}

\begin{equation}
V_+ = \frac{4}{3}\pi r^3_+.
\label{eqn:v}
\end{equation}
For the values of mass, $M$ (\ref{eqn:m}) and the temperature, $T_+$ (\ref{eqn:t}), using the first law of thermodynamics (\ref{fst}) of BH, we can compute the entropy of the black hole  from the relation, $S_+=\int \frac{1}{T_+}\left( \frac{\partial M}{\partial r_+}\right)dr_+$ as
 \begin{equation}
S_+=\pi  \left[\frac{e^{k/r_+}}{k} \left(k r_+ (k+r_+)-2 g^3\right)-k^2 \text{Ei}\left(\frac{k}{r_+}\right)\right], 
\label{eqn:s}
\end{equation}
where ``Ei" is the exponential integral function. Here, we find that entropy does not depend on the CS parameters $(a)$. Here, entropy does not follow the usual entropy area law of a black hole, $S_+=\frac{A}{4}$, (where $A=4 \pi r_{+}^2$ is the area of the event horizon) in the presence of a magnetic charge, $g$ and the deviation parameter, $k$. However, in the limit of parameters $g$ and $k$ tends to zero, we obtain the usual entropy area law as
\begin{equation}
\label{nledent0}
 S_+^{Swzc} = \pi r_{+}^2=\frac{A}{4},
\end{equation}
which follows the standard Bekenstein-Hawking area law and exactly matches the entropy of the four-dimensional Schwarzschild black hole. 
 
  Now, we also checked the expression for Hawking temperature evaluated from the first law of thermodynamics (\ref{fst}) as $T_H=\left(\ \frac {\partial M}{\partial S}\right)$ and found to agree with the one obtained in (\ref{eqn:t}) for a fixed value of magnetic charge $(dg=0)$. Hence, our solution for this black hole follows the first law of thermodynamics for the fixed value of the magnetic charge.
 
 It has been demonstrated by Wald \cite{wald1993black} that the entropy of a black hole obeys the area law, but in the case of regular black holes, one doesn't get the usual area form using the first law of thermodynamics. The deviation of the entropy (\ref{eqn:s}) relies on the general structure of the EMT of matter fields for regular black holes. { The mass term is modified in the presence of NLED with an extra factor. The modified mass is \cite{Rodrigues:2022zph,Rodrigues:2022qdp}
\begin{equation}
d {\cal M}=\left(1+4\pi \int_{r_+}^{\infty}r_+^2\frac{\partial T^t_t}{\partial M} dr_+\right)dM= \mathcal{C}(M,g,r_+)\,dM.
\end{equation}
The $\mathcal{C}(M,g,r_+)$ is the correction term. The modified first law of black hole thermodynamics is
\begin{equation}
d{\cal M}=T_+dS_+ + \Phi_{g} dg + V_+dP_+,
\end{equation}
the conventional form of the first law gets modified with an extra factor \cite{ma2014corrected}
\begin{equation}
 \mathcal{C}(M,g,r_+)\,dM=T_+ \,dS+ \Phi_{g} dg + V_+dP_+,\label{cor}
\end{equation}
where $T_+$
 is the Hawking temperature and $\mathcal{C}(M,g,r_+)$ is
\begin{equation}
 \mathcal{C}(M,g,r_+)=1+4\pi \int_{r_+}^{\infty}r_+^2\frac{\partial T^t_t}{\partial M} dr_+.
\end{equation}

}
We recover the conventional form of the first law of black hole thermodynamics when the factor $\mathcal{C}(M,r_+)$=1, as the EMT does not depend upon mass. Since any black hole
has temperature, it can be seen as a thermodynamic system. Thus, the conventional thermodynamic laws must be satisfied. We have two choices to connect Eq. (\ref{cor}) with the first law of thermodynamics. We know that $\delta E =T \delta S$  then the $E\to M$ and the  entropy becomes 
\begin{equation}
 \delta S_+= \mathcal{C}(M,g,r_+)\delta M.
 \label{mlaw}
\end{equation}
Following this modified first law of thermodynamics (\ref{mlaw}), the obtained entropy (\ref{eqn:s}) of this regular black hole solution follows the usual area law (\ref{nledent0}) of black hole mechanics.

\begin{figure}[!]
\begin{tabular}{c c} 
\includegraphics[width=.50\linewidth]{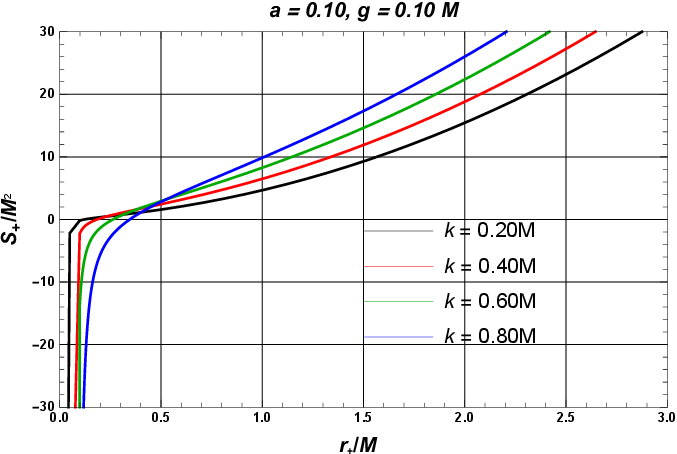} 
\includegraphics[width=.50\linewidth]{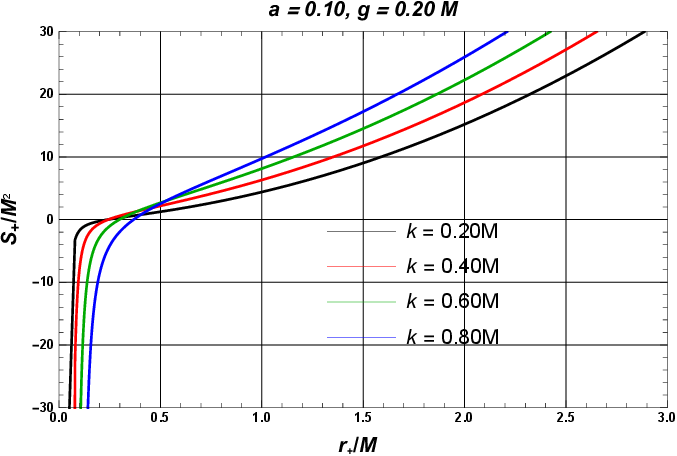}\\
\includegraphics[width=.50\linewidth]{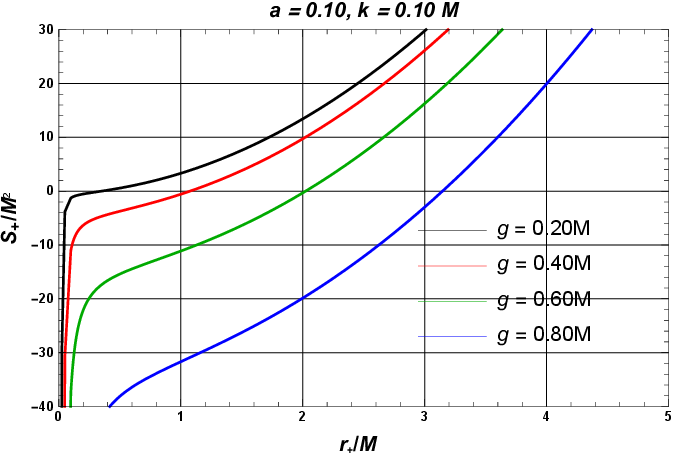}
\includegraphics[width=.50\linewidth]{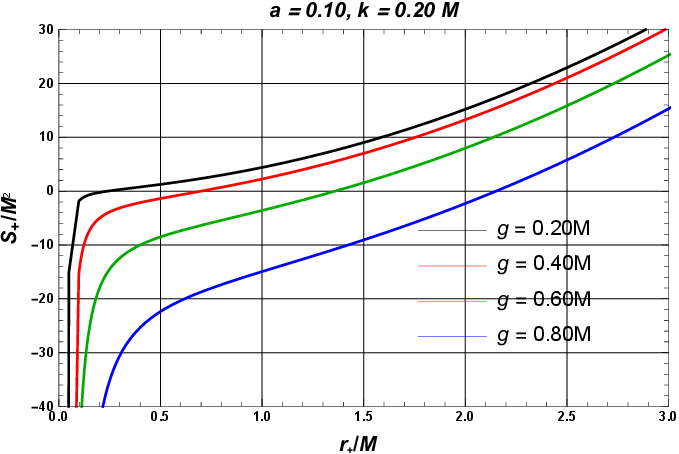}
\end{tabular}
\caption[Entropy plot of $4D$ NLED black hole in a CS]{Plot of entropy, $S_+$ versus horizon radius, $r_+$ of $4D$ NLED black hole in a CS for different value of $k$ and $g$.}
\label{fig:s} 
\end{figure}

To analyse the behaviour of entropy (\ref{eqn:s}) with the horizon radius of this $ 4$D ad black hole with NLED sources in the CS, we plot it in Fig. \ref{fig:s} for different values of  $g$ and   $k$. From the plot, we see that the entropy of a black hole increases with the value of $k$ while it has the opposite variation for $g$. Further, we observe that entropy is collinear in the region of this black hole's large horizon radius compared to that of the Schwarzschild black hole.

 Now, to understand the thermodynamic stability of this black hole, we first study the nature of its specific heat capacity $(C_+)$ to appreciate its local stability since the negative and positive signatures of the heat capacity define the unstable and the stable thermodynamical system of the black hole, respectively.

 The specific heat capacity $C_+$ for the black hole can be obtained from the relation $C_+=\left(\frac{\partial M}{\partial T_+}\right)=\left(\frac{\partial M}{\partial r_{+}}\right)\left(\frac{\partial r_+}{\partial T_+}\right)$, by using the expression for temperature and entropy from the equations. (\ref{eqn:t}) and (\ref{eqn:s}), respectively as
\begin{equation}
\label{eqn:c}
C_+=\frac{2 \pi  \left(g^3+r_+^3\right)^2\left[g^3 \left(k \left(\xi -r_+^2\right)+2 \xi  r_+\right)+r_+^3 \left(k \left(\xi -r_+^2\right)+3 r_+^3-\xi  r_+\right)\right]e^{k/r_+} }{2 g^3 r_+^4 \left(6 r_+^3-5 \xi  r_+ -2 k \xi \right)+r_+^7 \left(3 r_+^3+\xi  r_+ -2 k \xi\right)-2 g^6 \xi  r_+ (k+r_+)},
 \end{equation} 
where $\xi =(a-1) l^2$.
 
  From the expression of the specific heat capacity (\ref{eqn:c}), it is clear that for the $4D$ AdS black hole with NLED sources in the CS, the heat capacity was found to depend on all parameters $a$, $g$, $k$ and $l$. In the limit of $a$, $g$ and $k$ tends to zero, we obtain the expression of the heat capacity as
\begin{equation}
C_+^{Swzc} = 2 \pi  r_+^2 \left(\frac{3r_+^2+l^2}{3 r_+^2-l^2}\right),
\end{equation}
which is the expression for the heat capacity of the $4D$ Schwarzschild black hole in $AdS$ spacetime. { The heat capacity of the obtained black hole solution (\ref{bhsnled})  reduces to the heat capacity of the Letelier black hole with CS in the absence of magnetic monopole charge, $AdS$ Hayward black hole in the absence of ($k,a$) \cite{Luo:2023ndw,Sadeghi:2024krq} and $AdS$ regular black hole in the absence of ($g,a$) \cite{Singh:2022xgi}.}

\begin{figure}[!]
\begin{tabular}{c c} 
\includegraphics[width=.50\linewidth]{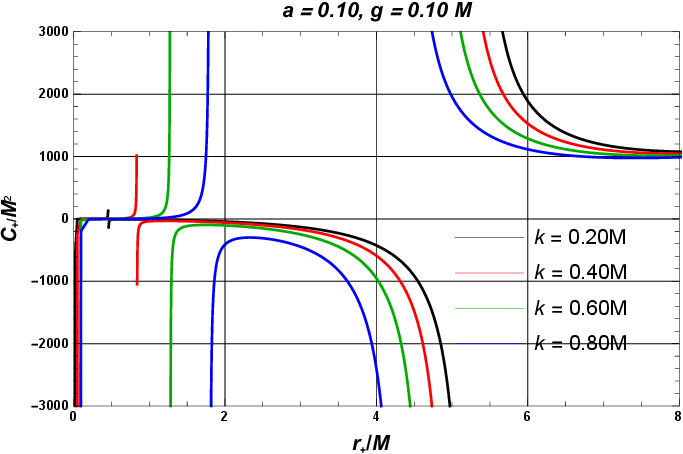} 
\includegraphics[width=.50\linewidth]{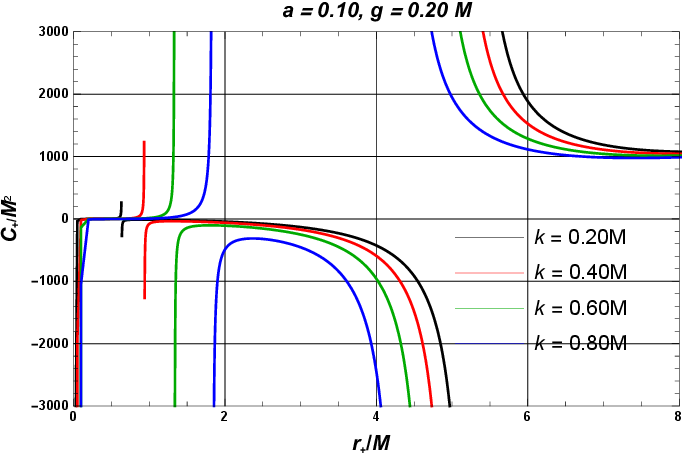}\\
\includegraphics[width=.50\linewidth]{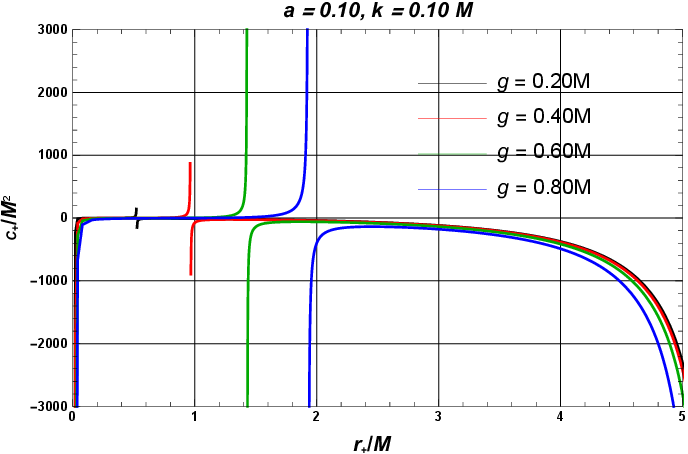}
\includegraphics[width=.50\linewidth]{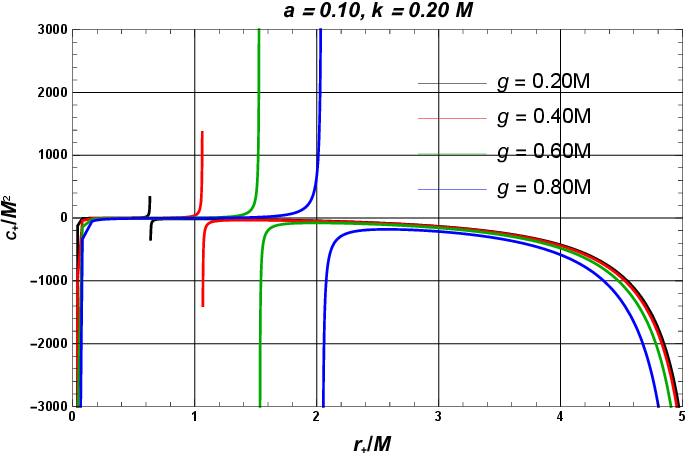}
\end{tabular}
\caption{The Plot of  heat capacity, $C_+$ versus horizon radius, $r_+$ of $4D$ NLED black hole in a CS for different value of $k$, $g$ and $a$ with $l=10$.}
\label{fig:c} 
\end{figure}

It is clumsy to point out the signature of the heat capacity of this regular black hole from its expression. Hence, we plot the heat capacity with the horizon radius in Fig. \ref{fig:c} by fixing the parameters $a$, $g$ and $k$. For different values of parameters, the heat capacity in all the cases shows asymptotic behaviour and is found to be discontinuous at a particular horizon radius, say at $r_+=r_{h1}$ and $r_+=r_{h2}$. For the region $r_+<r_{h1}$ and $r_+>r_{h2}$, the specific heat is found to have a positive value, i.e. $C_+ >0$ and hence the black hole is thermodynamically stable. In contrast, it is thermodynamically unstable for the region $r_{h1}<r_+<r_{h2}$ as specific heat has a negative value, i.e. $C_+ <0$. The divergence of heat capacity at horizon radius $r_+=r_{h1}$ and $r_+=r_{h2}$ resembles the second-order phase transition between stable and unstable phases of the black hole when it changes size from more minor to larger or vice versa.

 For further analysis, we have numerically computed the values of the particular horizon radius $r_{h1}$ and $r_{h2}$ along with the horizon region between these (unstable black hole region), $\Delta =(r_{h2}-r_{h1})$ as shown in Table \ref{tab:shc} for different values of magnetic charge, $g$, deviation parameter, $k$ and the CS parameters, $a$. As the values of these parameters increase, the particular horizon radius $r_{h1}$ and $r_{h2}$ increases and decreases, respectively. Consequently, the region between this horizon radius, $\Delta$, decreases. Hence, as the value of black hole parameters $a, g$ and $k$ increases, the unstable region (negative heat capacity) of a black hole, $\Delta$, reduces, i.e. the stability of a black hole increases for NLED sources in the presence of a CS.
 
\begin{table}[h!tb]
\centering 
\begin{tabular}{cccccccc}
\hline
\multicolumn{8}{c}{$a   = 0.10$}                                                                                                                                                                \\ \hline \hline
\multicolumn{4}{c|}{$k = 0.10$}                                                                     & \multicolumn{4}{c}{$g = 0.10$}                                                             \\ \hline
$g$                  & $r_{h1}$             & $r_{h2}$             & \multicolumn{1}{c|}{$\Delta$} & $k$                  & $r_{h1}$             & $r_{h2}$             & $\Delta$             \\ \hline
0.20                  & 0.524                & 5.372                & \multicolumn{1}{c|}{4.849}    & 0.20                  & 0.459                & 5.265                & 4.806                \\
0.40                  & 0.965                & 5.358                & \multicolumn{1}{c|}{4.393}    & 0.40                  & 0.837                & 5.022                & 4.185                \\
0.60                  & 1.430                & 5.317                & \multicolumn{1}{c|}{3.888}    & 0.60                  & 1.277                & 4.732                & 3.455                \\
0.80                  & 1.931                & 5.233                & \multicolumn{1}{c|}{3.302}    & 0.80                  & 1.797                & 4.356                & 2.559                \\ \hline \hline
\multicolumn{1}{l}{} & \multicolumn{1}{l}{} & \multicolumn{1}{l}{} & \multicolumn{1}{l}{}          & \multicolumn{1}{l}{} & \multicolumn{1}{l}{} & \multicolumn{1}{l}{} & \multicolumn{1}{l}{} \\ \hline
\multicolumn{8}{c}{$a   = 0.20$}                                                                                                                                                                \\ \hline \hline
\multicolumn{4}{c|}{$k = 0.10$}                                                                     & \multicolumn{4}{c}{$g = 0.10$}                                                             \\ \hline
$g$                  & $r_{h1}$             & $r_{h2}$             & \multicolumn{1}{c|}{$\Delta$} & $k$                  & $r_{h1}$             & $r_{h2}$             & $\Delta$             \\ \hline
0.20                  & 0.524                & 5.059                & \multicolumn{1}{c|}{4.535}    & 0.20                  & 0.460                & 4.951                & 4.491                \\
0.40                  & 0.967                & 5.042                & \multicolumn{1}{c|}{4.075}    & 0.40                  & 0.839                & 4.704                & 3.865                \\
0.60                  & 1.436                & 4.996                & \multicolumn{1}{c|}{3.560}    & 0.60                  & 1.287                & 4.404                & 3.117                \\
0.80                  & 1.949                & 4.898                & \multicolumn{1}{c|}{2.948}    & 0.80                  & 1.836                & 3.999                & 2.163                \\ \hline \hline
\end{tabular} 
\caption[Numerical data of stability and unstability regions of $4D$ NLED black hole in CS for different values of parameters $a, g$ and $k$]{For different values of parameters $a, g$ and $k$ with $l=10$, particular horizon radius $(r_{h1})$ and $(r_{h2})$ along with unstable region, $\Delta$ has been tabulated numerically for $4D$ NLED black hole in CS.}
\label{tab:shc}
\end{table}


 Further, to understand the global thermal stability of the black hole system, Gibbs free energy, $G_+$, plays one of the crucial roles in a thermodynamically stable system of the black hole $G_+\leq 0$. The Gibbs free energy of the black hole can be computed from the standard relation, $G_+ = M - T_+S_+$ by using the expression of mass, temperature and entropy from equations (\ref{eqn:m}), (\ref{eqn:t}) and (\ref{eqn:s}), respectively as

\begin{equation}
 G_+=\frac{{\left[ k \left(r_+^3+ g^3\right)  \left(r_+^2-\xi \right)e^{k/r_+}\right]}{-\left[\left(\frac{r_+^3-2 g^3}{r_+^3+g^3}\right)r_+ +2 r_+^3-k\right] \left[\left(k^2 r_+ +k r_+^2 -2 g^3\right)e^{k/r_+}-k^3 \text{Ei}\left(\frac{k}{r_+}\right)\right]}}{4 l^2 r_+^2}.
\label{eqn:g}
\end{equation}
In the limit of parameters $a$, $g$, and $k$ tends to zero, we obtain the Gibbs free energy as
\begin{equation}
G_+^{Swzc} = \frac{1}{4} \left(1-\frac{r_+^2}{l^2}\right) r_+,
\end{equation}
which is the expression of Gibbs free energy for $4D$ Schwarzschild black hole in $AdS$ spacetime. { The Gibbs free energy of the obtained black hole solution (\ref{bhsnled})  reduces to the Gibbs free energy of the Letelier black hole with CS in the absence of magnetic monopole charge, $AdS$ Hayward black hole in the absence of ($k,a$) \cite{Luo:2023ndw,Sadeghi:2024krq} and $AdS$ regular black hole in the absence of ($g,a$) \cite{Singh:2022xgi}.}

\begin{figure}[!]
\begin{tabular}{c c} 
\includegraphics[width=.50\linewidth]{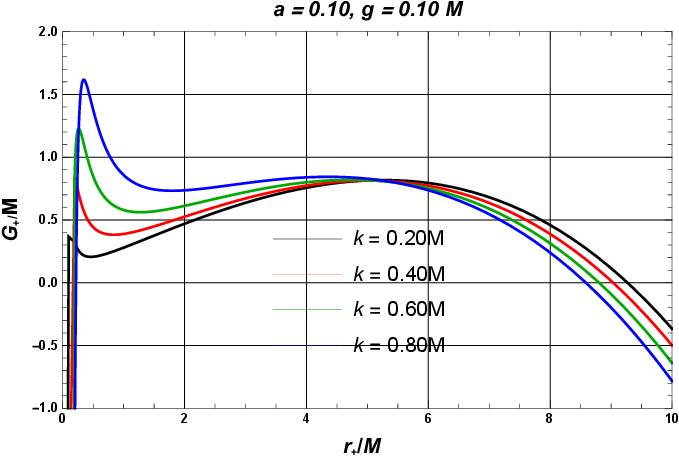} 
\includegraphics[width=.50\linewidth]{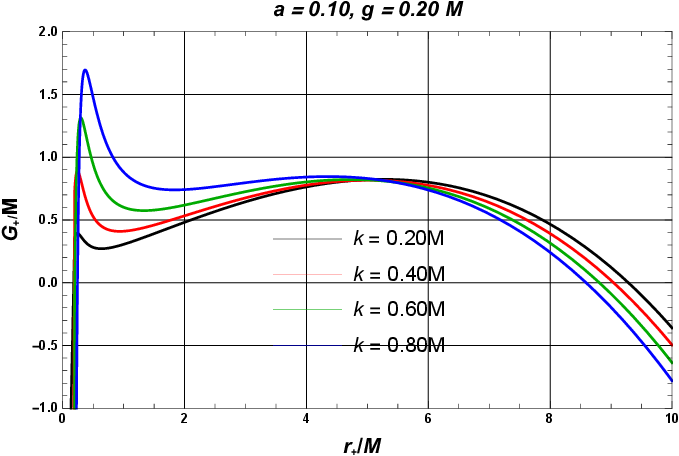}\\
\includegraphics[width=.50\linewidth]{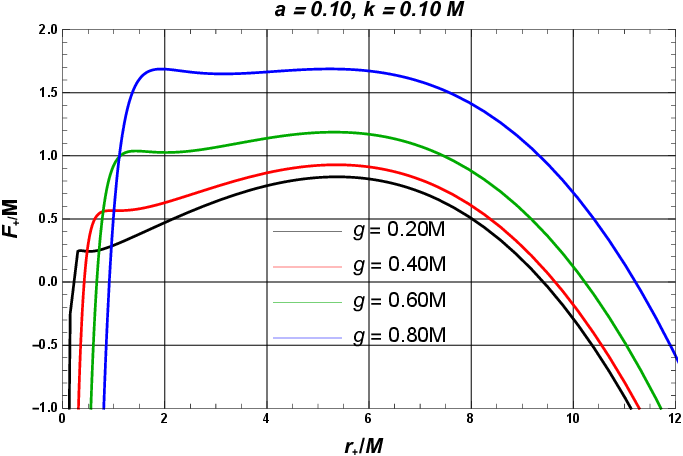}
\includegraphics[width=.50\linewidth]{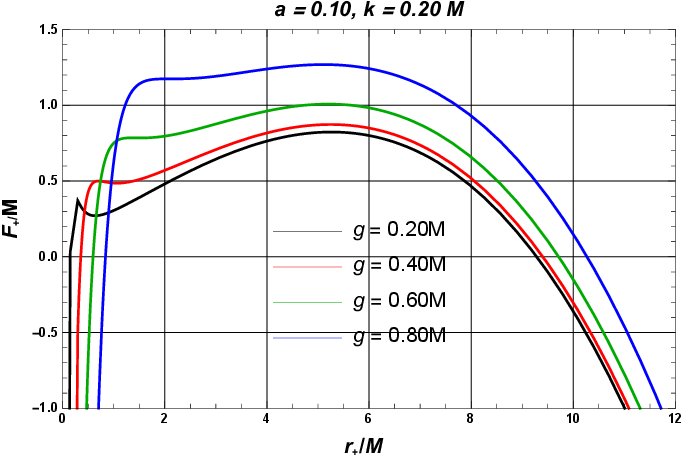}
\end{tabular}
\caption[Gibbs free energy plot of $4D$ NLED black hole in a CS]{Plot of Gibbs free energy, $G_+$ versus horizon radius, $r_+$ of $4D$ NLED black hole in a CS for different value of $k$, $g$ and $a$ with $l=10$.}
\label{fig:g} 
\end{figure}

For the analysis of the global stability of the $4D$ AdS black hole with NLED sources in the presence of a CS, Gibbs free energy is plotted in Fig. \ref{fig:g} with horizon radius for different values of a CS parameter $(a)$, $g$ and $k$.

From the plot Fig. \ref{fig:g}, we see that the effect of a CS parameter $a$, $g$ or  $k$ is more significant for small black holes. In the variation of $g$ or $k$ for a fixed value of $a$, Gibbs free energy attains a maximum positive value $(G_+^{max} > 0)$ and then it drops to a local minimum value, then increases to a local maximum value and remains positive, after that it attains the local maximum value, from where its slope turns negative and Gibbs free energy becomes negative to coincide with the $4D$ AdS Schwarzschild black hole in large black hole region.

\begin{figure}[h]
 \centering
\includegraphics[scale=0.8]{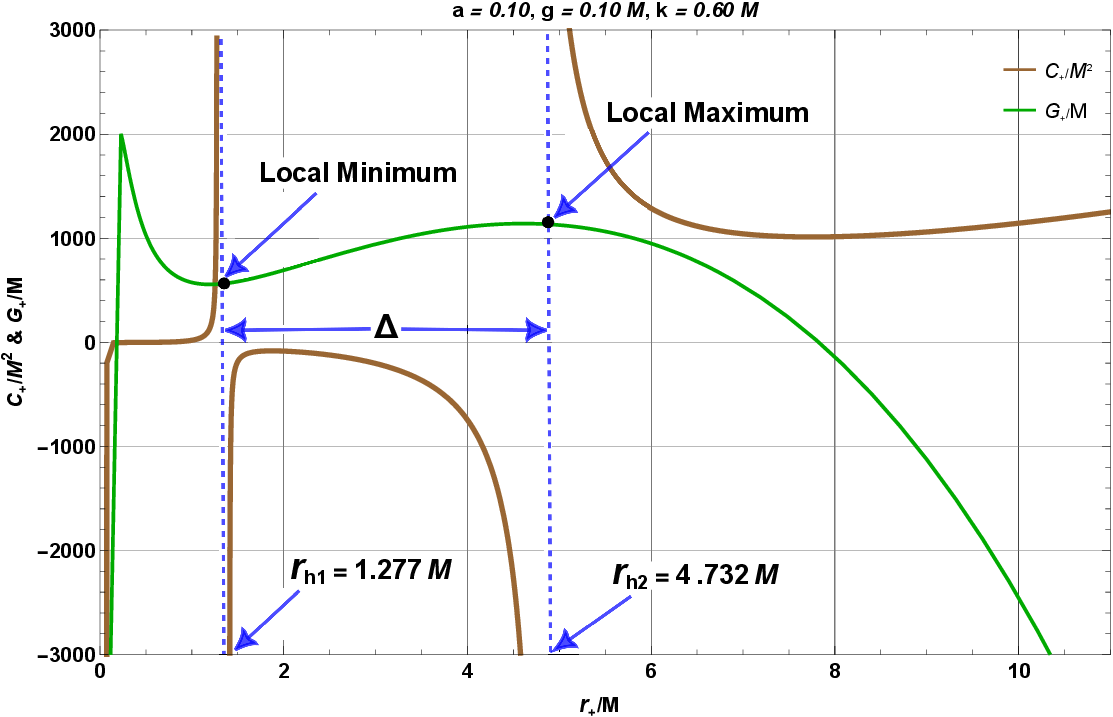} 
\caption[Depiction of Hawking-Page phase transition of $4D$ NLED black hole in a CS]{Depiction of Hawking-Page phase transition of $4D$ NLED black hole in a CS for $a=0.10, g=0.10 M$ and $k=0.60 M$ with $l=10$, showing local minimum and local maximum at horizon radius  $r_+=r_{h1}$ and $r_+=r_{h2}$, respectively (Graph not to scale in vertical axis).}
\label{hpt} 
\end{figure}

  Here, we observe that the Gibbs free energy attains local minimum and local maximum values at the particular horizon radius, $r_{h1}$ and $r_{h2}$, respectively, at which the specific heat capacity (\ref{eqn:c}) flip its sign (see Table \ref{tab:shc} and Fig. \ref{fig:c}).
For the horizon radius, $r_+>r_{h1}$, the slope of the Gibbs free energy has an increasing function (positive slope) with the horizon radius and achieves the local maximum value at the horizon radius $r_+=r_{h2}$, after which its slope turns negative. Hence, it provides the theory of the usual Hawking-Page phase transition of the black hole. For a particular case of $a=0.10, g=0.10$ and $k=0.60$, these have been depicted in Fig. \ref{hpt}. Further, it is observed that as the values of parameters increase, $a, g$ and $k$ increase, the horizon region between the local minimum and local maximum decreases, and hence, similar to the case of specific heat capacity, the unstable region of a black hole, $\Delta$ reduces and therefore stability region of this NLED black hole in the presence of CS increase.
 
\section{$P-v$ Criticality of NLED Black Hole in CS}\label{sec6.4}
 
In this section, we study the critical behaviour and phase transition of the $4D$ AdS black hole solution with NLED sources in the presence of CS in the extended phase space analogous to the VdW fluid. Since in the extended thermodynamics, the negative cosmological constant $(\Lambda <0)$ induces the thermodynamic pressure of the black hole as shown in equation (\ref{eqn:p}) and the thermodynamic volume (\ref{eqn:v}) is conjugate to the pressure, hence it can be interpreted as the change in mass of the black hole under the variation of the pressure having fixed horizon area as $V_+=\left(\frac{\partial M}{\partial P_+}\right)_{S_+}$. The mass of the black hole $(M)$ is then equivalent to the enthalpy of the black hole system.

 The equation of the state for this black hole thermodynamic system can be obtained by putting the value of $l$  from equation (\ref{eqn:p}) in equation (\ref{eqn:t}) as
 
 \begin{equation}
P_+=\frac{3}{8 \pi  r_+^2}\left[\frac{r_+^3 (k-a k+r_+ (-1+a+4 \pi  r_+ T_+))+g^3 (k-a k+2 r_+ (1-a+2 \pi  r_+ T_+))}{3r_+^4- k\left(r_+^3+g^3  \right)}\right],
 \label{eqn:eos}
 \end{equation}
and comparing the equation (\ref{eqn:eos}) with the VdW equation of the fluid \cite{goldenfeld2018lectures}
 \begin{equation}
\label{eqn:vdw}
P=\frac{T}{v-\beta}-\frac{\alpha}{v^2}\approx\frac{T}{v}+\frac{\beta T-\alpha}{v^2}+O(v^{-3}),
\end{equation}
where $v$ is the specific volume of the fluid and constants $\alpha$ and $\beta$ (both $>0$) represent the measure of the attraction between molecules and the size of the molecules in the fluid, respectively. We obtain the specific volume of this black hole thermodynamics system as \cite{Sudhanshu:2023zle}

\begin{equation}
\label{eqn:sv}
v=2 \left(\frac{r_+^4}{r_+^3 + g^3}-\frac{k}{3}\right).
\end{equation}
In the limit $g$ and $k$ tend to zero, the specific volume becomes $v=2r_+$.

 The critical components of temperature, $T_c$ pressure, $P_c$ and specific volume, $v_c$  or horizon radius, $r_c$, can be determined from the condition of inflection point at critical components  $P_c$ and $v_c$ by satisfying the following conditions
 \begin{equation}
 \label{eqn:cond}
 \left(\frac{\partial P_+}{\partial r_+}\right)=0  \quad \text{and}\quad \left(\frac{\partial^2 P_+}{\partial r_+^2}\right)=0.
 \end{equation}
Using the critical point conditions (\ref{eqn:cond}) in the equation of state (\ref{eqn:eos}), one gets the following equation.
 
 \begin{equation}
 \label{eqn:rc}
\frac{3 (a-1) \left(4 g^6 (6 k+5 r_+)+g^3 r_+^3 (9 k+28 r_+)+r_+^6 (3 k-r_+)\right)}{4 \pi  r_+^4 \left(4 g^3+r_+^3\right) \left(g^3 k+r_+^3 (k-3 r_+)\right)}=0
 \end{equation}
 The critical point horizon radius, $r_c$, can be obtained by solving the equation (\ref{eqn:rc}). However, one can't solve it analytically, but
  the critical radius $r_c$ (or specific volume from equation (\ref{eqn:sv})), critical temperature $T_c$ and critical pressure $P_c$ can be computed numerically for the specific values of   parameters $a$, $g$ and $k$ of this $4D$ AdS black hole with NLED sources in the CS as shown in Table  
  \ref{tab:cc}. The universal ratio, critical compressibility factor, $Z_c$, which describes the deviation of the real thermodynamic system from its behaviour as an ideal gas, is calculated by relation  $Z_c=\frac{P_c v_c}{T_c}$.

\begin{table}[h!tb]
 \centering
\begin{tabular}{ccccclccccc}
\cline{1-5} \cline{7-11}
\multicolumn{5}{c}{$a = 0.10$, $g = 0.10$}                                                                             & \multicolumn{1}{c}{} & \multicolumn{5}{c}{$a = 0.20$, $g = 0.10$}                                                                             \\ \cline{1-5} \cline{7-11} 
\multicolumn{1}{c|}{$k$} & $r_c$                & $P_c$                & $T_c$                & $Z_c$                &                      & \multicolumn{1}{c|}{$k$} & $r_c$                & $P_c$                & $T_c$                & $Z_c$                \\ \cline{1-5} \cline{7-11} 
\multicolumn{1}{c|}{0.20} & 0.669                & 0.029                & 0.109                & 0.318                &                      & \multicolumn{1}{c|}{0.20} & 0.669                & 0.026                & 0.096                & 0.318                \\
\multicolumn{1}{c|}{0.40} & 1.221                & 0.008                & 0.057                & 0.310                &                      & \multicolumn{1}{c|}{0.40} & 1.221                & 0.007                & 0.051                & 0.310                \\
\multicolumn{1}{c|}{0.60} & 1.809                & 0.004                & 0.038                & 0.308                &                      & \multicolumn{1}{c|}{0.60} & 1.809                & 0.003                & 0.034                & 0.308                \\
\multicolumn{1}{c|}{0.80} & 2.405                & 0.002                & 0.029                & 0.308                &                      & \multicolumn{1}{c|}{0.80} & 2.405                & 0.002                & 0.026                & 0.308                \\ \cline{1-5} \cline{7-11} 
\multicolumn{1}{l}{}     & \multicolumn{1}{l}{} & \multicolumn{1}{l}{} & \multicolumn{1}{l}{} & \multicolumn{1}{l}{} &                      & \multicolumn{1}{l}{}     & \multicolumn{1}{l}{} & \multicolumn{1}{l}{} & \multicolumn{1}{l}{} & \multicolumn{1}{l}{} \\ \cline{1-5} \cline{7-11} 
\multicolumn{5}{c}{$a = 0.10$, $k = 0.10$}                                                                             &                      & \multicolumn{5}{c}{$a = 0.20$, $k = 0.10$}                                                                             \\ \cline{1-5} \cline{7-11} 
\multicolumn{1}{c|}{$g$} & $r_c$                & $P_c$                & $T_c$                & $Z_c$                &                      & \multicolumn{1}{c|}{$g$} & $r_c$                & $P_c$                & $T_c$                & $Z_c$                \\ \cline{1-5} \cline{7-11} 
\multicolumn{1}{c|}{0.20} & 0.737                & 0.032                & 0.122                & 0.357                &                      & \multicolumn{1}{c|}{0.20} & 0.737                & 0.028                & 0.108                & 0.357                \\
\multicolumn{1}{c|}{0.40} & 1.341                & 0.010                & 0.072                & 0.368                &                      & \multicolumn{1}{c|}{0.40} & 1.341                & 0.009                & 0.064                & 0.368                \\
\multicolumn{1}{c|}{0.60} & 1.951                & 0.005                & 0.051                & 0.372                &                      & \multicolumn{1}{c|}{0.60} & 1.951                & 0.004                & 0.045                & 0.372                \\
\multicolumn{1}{c|}{0.80} & 2.562                & 0.003                & 0.039                & 0.374                &                      & \multicolumn{1}{c|}{0.80} & 2.562                & 0.003                & 0.035                & 0.374                \\ \cline{1-5} \cline{7-11} 
\end{tabular}
\caption[Numerical data of critical temperature $T_c$, critical pressure $P_c$ and critical compressibility factor, $Z_c$ of $4D$ NLED black hole in CS for different values of parameters $a, g$ and $k$]{The critical temperature $T_c$, critical pressure $P_c$ and critical compressibility factor, $Z_c$ for different values of parameters $a$, $g$ and $k$ has been tabulated numerically for $4D$ NLED black hole with CS.}
\label{tab:cc} 
\end{table}

\begin{figure}[]
\centering 
\begin{tabular}{c c} 
\includegraphics[width=.50\linewidth]{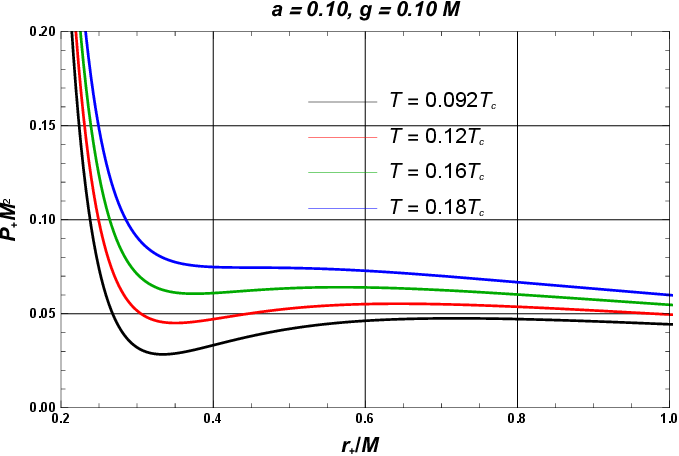} 
\includegraphics[width=.50\linewidth]{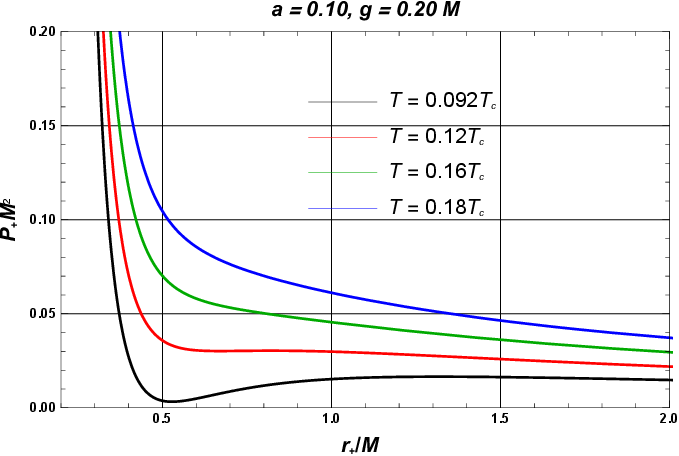}\\
\includegraphics[width=.50\linewidth]{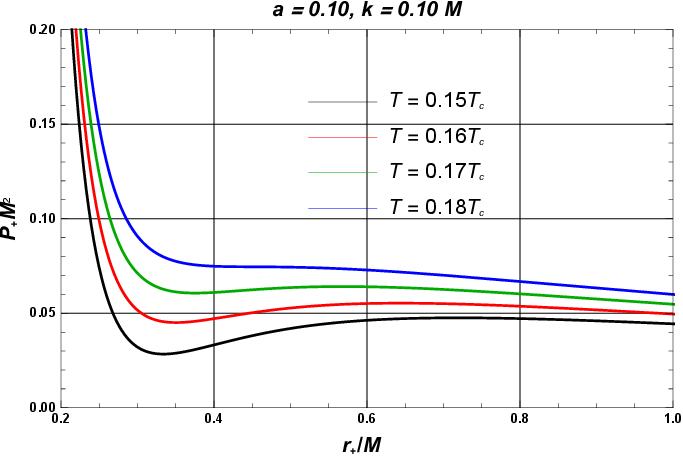}
\includegraphics[width=.50\linewidth]{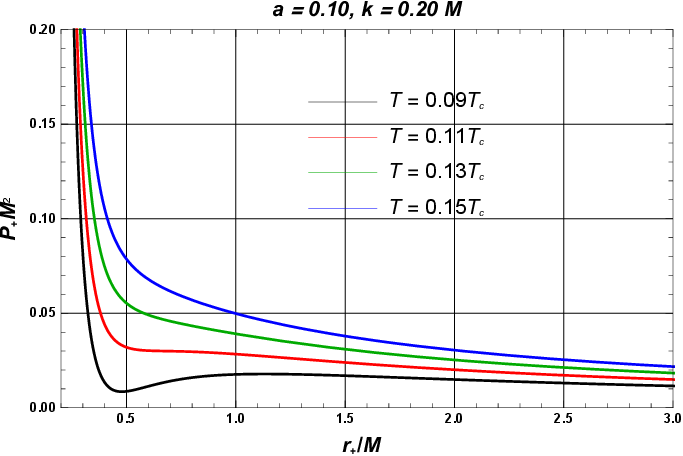}\\
\end{tabular} 
\caption[Pressure versus horizon radius plot at a critical temperature of $4D$ NLED black hole in a CS]{Plot of pressure, $P_+$ versus horizon radius, $r_+$ at the corresponding critical temperature, $T_c$ of $4D$ NLED black hole in a CS for different value of $k$, $g$ and $a$.}
\label{fig:cc} 
\end{figure}

 To analyse the behaviour of black hole pressure with the horizon radius at the isotherm, we plot the expression of pressure (\ref{eqn:p}) for different values of parameters $a$, $g$ and $k$ as shown in Fig. \ref{fig:cc}.

 From Table \ref{tab:cc} and Fig. \ref{fig:cc}, it is clear that the critical horizon radius, $r_c$ and hence critical specific volume, $v_c$, increases with the increase in the values of $g$ or $k$, but found to be independent of CS parameter $(a)$. Though the universal compressibility ratio, $Z_c$, increases with an increase in the value of the parameter magnetic charge, $g$ and decreases with an increase in the value of the deviation parameter, $k$, it is found to be independent of a CS parameter, $a$. The universal ratio, $Z_c$, is found to be analogous to that of the VdW fluid, for which its universally accepted value is $0.375$. { It means that the deviation parameter and magnetic charge behaviour are opposite on the $Z_c$. Both critical pressure and critical temperature decrease with increasing values of parameters $a$, $g$, and $k$. It is important to note that the critical radius (or critical specific volume) of the black hole thermodynamic system increases with the deviation parameter ($k$) and the magnetic charge ($g$). { Small black hole and large black hole are stable.}
However, intermediate black hole is unstable since the heat capacity C+ is negative (see Fig. 3).
When $T_+<T_{\star}$, the small black hole is obtained, and $T_+>T_{\star}$ corresponding to a large black
hole due to small free energy, where $T_{\star}$ is the transition temperature, which are 0.081, 0.042
for $k=0.2$ and $k=0.4$ with fixed value of $a = g = 0.1$, respectively. We can transit from one phase to another phase at critical temperature due to the same free energy.}
 
\begin{figure}[]
\begin{tabular}{c c} 
\includegraphics[width=.50\linewidth]{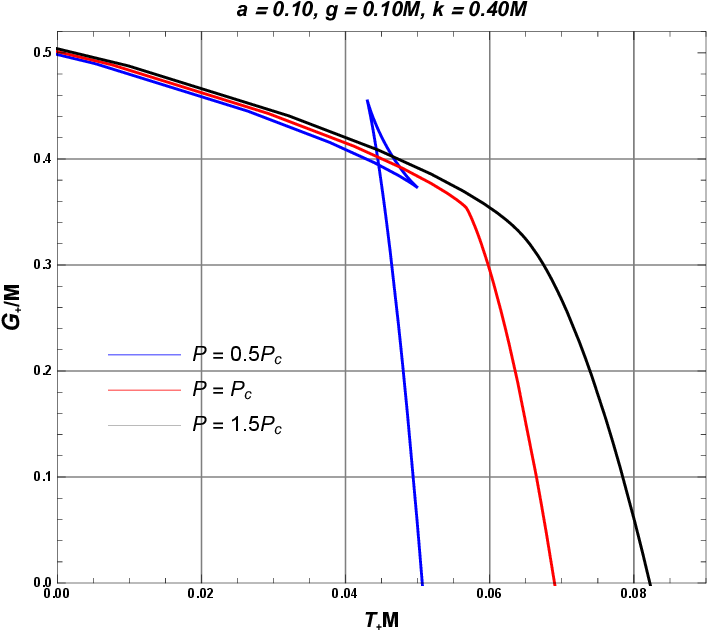}
\includegraphics[width=.50\linewidth]{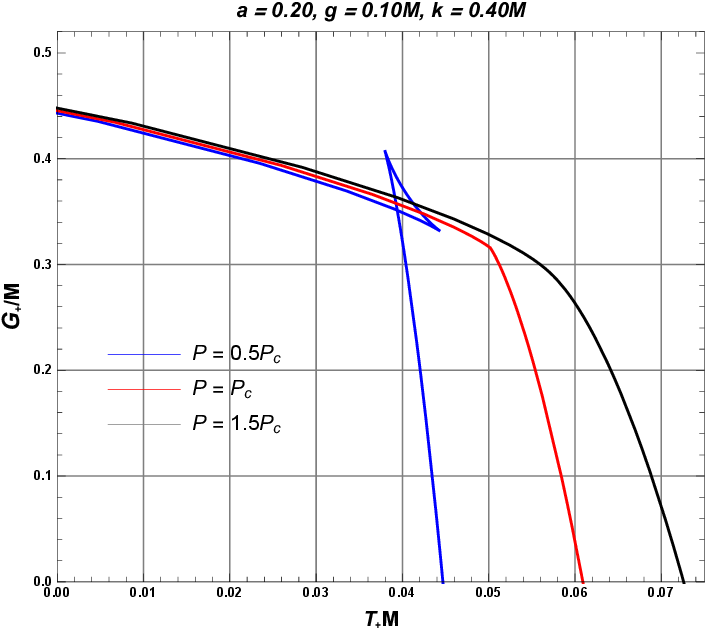}\\
\includegraphics[width=.50\linewidth]{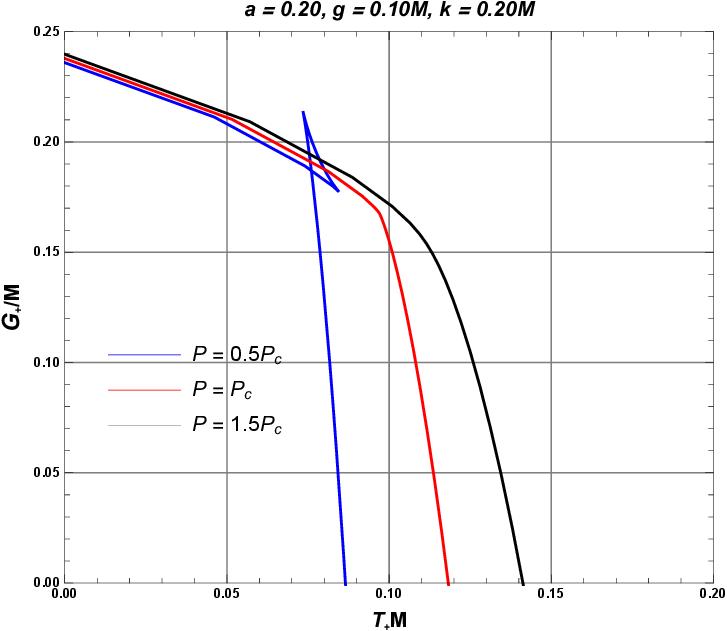}
\includegraphics[width=.50\linewidth]{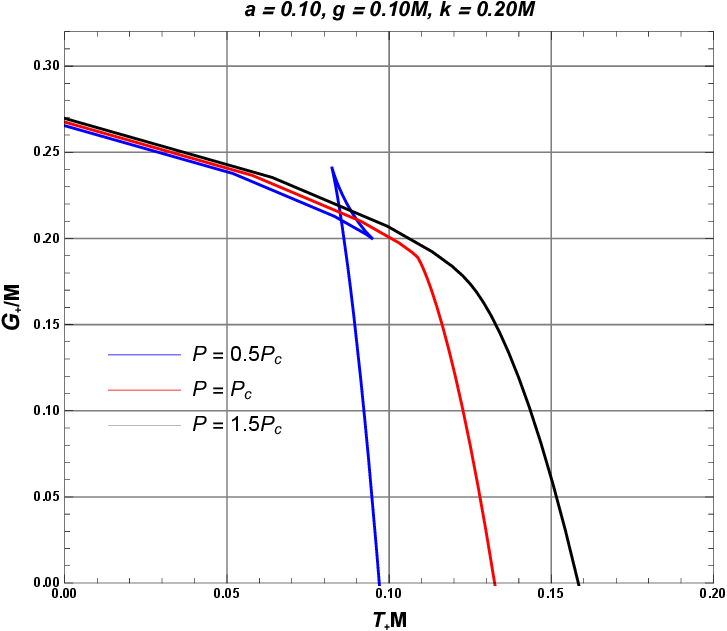}\\
\end{tabular}
\caption[Gibbs free energy versus temperature plot around the corresponding critical pressure of $4D$ NLED black hole in a CS]{Plot of Gibbs free energy, $G_+$ versus temperature, $T_+$ of $4D$ NLED black hole in a CS for different value of $k$, $g$ and $a$ around the corresponding critical pressure, $P_c$.}
\label{fig:gt} 
\end{figure}

 Now, to study the phase structure of the black hole thermodynamic system, we plot the Gibbs free energy (\ref{eqn:g}) versus temperature (\ref{eqn:t}) { for $P<P_c$, $P=P_c$ and $P>P_c$} with the variation of the magnetic charge  $(g)$, $k$ and the CS parameter $(a)$ of this black hole solution as shown in Fig. \ref{fig:gt}.  The Gibbs free energy, $G_+$, which analyses the phase transition of the black hole thermodynamic system analogous to the VdW fluid phase transition, shows variation with parameters $a$, $g$ and $k$ on the phase structure of the black hole system. {In $G_+-T_+$ plots, the appearance of characteristic swallow tail shows that the obtained values are critical for phase transition. In Fig. \ref{fig:gt}, we can see that the swallow tail shape exists when $P<P_c$ for the first order phase transition and $P=P_c$ for the second order phase transition. There is no phase transition when thermodynamic pressure is larger than the critical pressure $P_c$. The black hole transits from one phase to another due to the same free energy; the corresponding temperature is the transition temperature. The sub-critical isobar is the region where the phase transition takes place. It is noted that the Gibbs free energy of two phases is a decreasing function of CS parameter ($a$) and deviation parameter$(k)$.}

 In the $G_+$ vs $T_+$ plots in Fig. \ref{fig:gt}, the characteristic swallow tail indicates the critical values where the phase transition occurs. From the plot Fig. \ref{fig:gt}, we observe that the swallow tail exists for pressure, $P_+ < P_c$, which indicates the first order phase transition and the second order phase transition occurs for pressure, $P_+ = P_c$ in this black hole thermodynamic system.

\section{Results and Conclusions}\label{sec6.6}

 In this work, we have obtained a regular black hole solution in the context of general relativity coupled to NLED sources in the presence of CS in $4D$ AdS spacetime. The resulting solution is characterised by a cloud of string parameters, magnetic charge and deviation parameters, which at most have both Cauchy and event horizons. In the absence of these parameters, this solution corresponds to that of Schwarzschild black hole in AdS spacetime. We have investigated its thermodynamic properties both numerically and graphically. Its stability has been studied by observing heat capacity behaviour and Gibbs free energy. Ultimately, its $P-v$ criticality phenomena have been analysed based on the thermodynamics of the VdW fluid system. 
 
 The size of the black hole (event horizon) increases with the increase in the value of the cloud of string parameters. Still, it decreases with the increase in the value of magnetic charge or deviation parameters. Also, these parameters have critical values for which only the event horizon survives. Beyond this limit of parameters, no horizon has been found for this black hole, i.e., the existence of the black hole fades away. Furthermore, we have discussed this black hole's thermodynamics by deriving its mass's expression, Hawking temperature and entropy at the event horizon regarding black hole parameters. The mass of the black hole increases with the increase in values of the magnetic charge or deviation parameters. Still, it decreases with the increase in the value of the CS parameter. On the Hawking temperature, the effect of magnetic charge, deviation parameter or CS parameter is more significant for small black hole regions, where the temperature first increases to a maximum value for a particular horizon radius and then coincides with that of Schwarzschild black hole temperature in large black holes region.
Further, we observed that the maximum temperature decreases and shifts towards the area of an enormous horizon radius as the value of these parameters increases. We have noticed that the black hole solution follows the first law of thermodynamics for the fixed value of the magnetic charge. Hence, the black hole's entropy is independent of the cloud of string parameters and follows the usual entropy area law only in the modified first law of thermodynamics. Graphically, it has been observed that the entropy of the black hole has increased with the value of the deviation parameter while it has the opposite variation for magnetic charge.

  This black hole's local and global stability has been analysed by investigating the nature of heat capacity and Gibbs free energy, which is dependent on black hole parameters. The heat capacity flip sign at some particular horizon radius resembles the second-order phase transition between unstable and stable phases of the black hole when it changes size from more minor to larger or vice versa. As the values of magnetic charge, deviation parameter, and CS parameters increase, the unstable region of a black hole (negative heat capacity) reduces, allowing the black hole to be thermodynamically stable for NLED sources in the presence of a CS. The Gibbs free energy analysis has confirmed the existence of local minimum and local maximum where the heat capacity flips its sign. Hence, it provides the theory of the usual Hawking-Page phase transition of the black hole.
Further, it has been observed that as the values of parameters increase, the region between the local minimum and local maximum decreases; hence, similar to the case of specific heat capacity, the stability region of the NLED black hole increases in the presence of CS. The $P-v$ criticality of this black hole based on VdW fluid has also been studied. It was found that the critical values significantly depend on the string parameter, magnetic charge, and deviation parameters. The critical pressure and temperature decrease with the increase in the values of these parameters. It has been found that the critical horizon radius and, hence, specific volume increases with an increase in magnetic charge or deviation parameter. The universal compressibility ratio increases with an increase in the value of magnetic charge but decreases with an increase in the value of the deviation parameter. Both specific volume and universal compressibility ratio are independent of a CS parameter. The universal compressibility ratio is equivalent to that of the VdW fluid. Now, to glimpse into the phase structure of its thermodynamic system, the variation of Gibbs energy with temperature near critical pressure has been analysed graphically with the variation of the magnetic charge,  deviation parameter and CS parameter. Here, we have observed a characteristic swallow tail for pressure less than its critical value, which indicates the first-order phase transition. After the critical pressure of this black hole thermodynamic system, a second-order phase transition has been observed. 
  
  The main achievement of this work is to show that the thermodynamic properties and stability nature of black holes are affected by the parameters of NLED sources and CS in modified gravity. It would be exciting and challenging to focus on the properties of this black hole on higher dimensional solutions. A straightforward extension of our work is to study the black hole shadow problem, which becomes vital after its observational evidence, and the effects of correction due to thermal fluctuations, black bounces, etc, for these black hole solutions. It would also be interesting to analyse the properties of these black holes coupled with the sources of different forms of modified gravity like $f(R)$ gravity, $f(R, \phi)$ gravity, $f(T)$ gravity, $f(\mathcal{G})$ gravity,  Einstein-Gauss-Bonnet gravity, Yang-Mills fields, Ho\v{r}ava-Lifshitz gravity and so on. These will be the subject of future investigation.
\section*{Acknowledgement}
 This research was funded by the Science Committee of the Ministry of Science and Higher Education of the Republic of Kazakhstan (Grant No. AP22682760). DVS would like to thank DST-SERB for project no. EEQ/2022/000824.
 
\renewcommand{\bibname}{References}
\bibliographystyle{ieeetr} 
\bibliography{bibliography.bib}

\end{document}